\documentclass{aa}
\usepackage{natbib}
\bibpunct{(}{)}{;}{a}{}{,} 
\usepackage[varg]{txfonts}
\usepackage{xcolor}
\usepackage{array}

\begin{document}

\title{Hot WHIM counterparts of FUV \ion{O}{VI} absorbers: \\
Evidence in the line-of-sight towards quasar 3C~273}
\titlerunning{Hot WHIM counterparts of FUV \ion{O}{VI}: Evidence towards the 3C~273}

\author{Jussi Ahoranta\inst{1,2}
	\and Jukka Nevalainen\inst{3} 
	\and Nastasha Wijers\inst{4}
	\and Alexis Finoguenov\inst{1} 
	\and Massimiliano Bonamente\inst{2,5}
	\and Elmo Tempel\inst{3,6} 
	\and Evan Tilton\inst{7} 
	\and Joop Schaye\inst{4}
	\and Jelle Kaastra\inst{4,8}
	\and Ghassem Gozaliasl\inst{9,1}
	}

\institute{Department of Physics, University of Helsinki, P.O. Box 64, FI-00014, Finland
	\and Department of Physics, University of Alabama in Huntsville, Huntsville, AL, USA
	\and Tartu Observatory, University of Tartu, Observatooriumi 1, 61602 T\~{o}ravere, Estonia
	\and Leiden Observatory, Leiden University, PO Box 9513, NL-2300 RA Leiden, the Netherlands
	\and NASA National Space Science and Technology Center, Huntsville, AL, USA
	\and Leibniz-Institut f\"ur Astrophysik Potsdam (AIP), An der Sternwarte 16, D-14482 Potsdam, Germany
	\and Deparment of Physics \& Astronomy, Regis University, Denver, CO 80221, USA
	\and SRON Netherlands Institute for Space Research, Sorbonnelaan 2, 3584 CA Utrecht, The Netherlands
	\and Finnish centre for Astronomy with ESO (FINCA), Quantum, Vesilinnantie 5, University of Turku, FI-20014 Turku, Finland}

\date{Received / Accepted }

\abstract {} 
{We explore the high spectral resolution X-ray data towards the quasar 3C~273 to search for signals of hot ($\sim10^{6-7}$~K) X-ray-absorbing gas co-located with two established intergalactic FUV \ion{O}{VI} absorbers.}
{We analyze the soft X-ray band grating data of all XMM-\emph{Newton} and \emph{Chandra} instruments to search for the hot phase absorption lines at the FUV predicted redshifts. The viability of potential line detections is examined by adopting the constraints of a physically justified absorption model. The WHIM hypothesis is investigated with a complementary 3D galaxy distribution analysis and by detailed comparison of the measurement results to the WHIM properties in the \texttt{EAGLE} cosmological, hydrodynamical simulation.}
{At one of the examined FUV redshifts, $z=0.09017\pm0.00003$, we measured signals of two hot ion species, \ion{O}{VIII} and \ion{Ne}{IX}, with a $3.9\sigma$ combined significance level. While the absorption signal is only marginally detected in individual co-added spectra, considering the line features in all instruments collectively and assuming collisional equilibrium for absorbing gas, we were able to constrain the temperature ($kT=0.26\pm0.03$~keV) and the column density ($N_{\mathrm{H}}\times\mathrm{Z_\sun/Z}=1.3_{-0.5}^{+0.6}\times10^{19}$~cm$^{-2}$) of the absorber. Thermal analysis indicates that FUV and X-ray absorption relate to different phases, with estimated temperatures, $T_\mathrm{FUV}\approx3\times10^5$, and, $T_\mathrm{X-ray}\approx3\times10^6$~K. 
These temperatures match the \texttt{EAGLE} predictions for WHIM at the FUV/X-ray measured $N_\mathrm{ion}$-ranges.
We detected a large scale galactic filament crossing the sight-line at the redshift of the absorption, linking the absorption to this structure.}
{This study provides observational insights into co-existing warm and hot gas within a WHIM filament and estimates the ratio of the hot and warm phases.
Because the hot phase is thermally distinct from the \ion{O}{VI} gas, the estimated baryon content of the absorber is increased, 
conveying the promise of X-ray follow-up studies of FUV detected WHIM in refining the picture of the missing baryons.}

\maketitle

\section{Introduction}

The missing baryons problem (\citealt{persic92}) in the low-$z$ universe is likely a manifestation of the limitations of current observational capabilities at the high-energy end of the electromagnetic spectrum. The majority of the non-detected baryons are expected to reside in the warm-hot intergalactic medium (WHIM), the shock-heated diffuse gas accumulated within the large scale structures of dark matter in the Universe \citep{cen99,dave99}. Such concentrations of hot, highly-ionized gas are, in principle, observable in the FUV and X-ray bands. Whereas in the FUV band, measurements of broad Ly$\alpha$ and \ion{O}{VI} absorption lines in the local ($z<1$) universe have revealed the warm part ($T \sim 10^{5-6}$~K) of the WHIM (e.g., \citealt{tripp00}), the number of X-ray observations of the hot WHIM phase remains limited. The low expected column densities ($\lesssim10^{16}$~cm$^{-2}$) of the most prominent ions in the hot WHIM temperature range ($\sim10^{6-7}$ K), such as \ion{O}{VII-VIII}, \ion{Ne}{IX-X} and \ion{N}{VII} \citep[e.g.,][]{fang02,wijers18}, significantly limit the possibilities for studying the hot WHIM with observations using the present instrumentation. 

Currently, the method with the most potential for the direct detection of the WHIM uses quasar or blazar emission as a backlight against which absorption signatures might be found. In the FUV band this approach has been successfully applied \citep[e.g.,][]{tripp00, sembach01, williger10, tilton12, shull12, danforth16}, in contrast to the X-ray band where the observational evidence is typically less solid and therefore subject to other interpretations (see, e.g., discussions in \citealt{rasmussen07}, \citealt{max17}). Present high-energy resolution soft X-ray band instruments' (the XMM-\emph{Newton} RGS and \emph{Chandra} LETG grating spectrometers) effective areas vary from a few to few tens of cm$^2$ at the relevant energies, and consequently $\sim$ Ms exposure times are typically required to attain sufficient photon statistics for possible WHIM line detections. In such observations, systematics such as instrument calibration uncertainties require special attention.
In recent years, numerous studies of hot WHIM absorbers have been published, for example by \cite{buote09}, \cite{nicastro10}, \cite{ren14}, \cite{max16}, \cite{nicastro18}, and \cite{kovacs19}, among others. However, the number and the significance of the possible hot WHIM detections has remained low, and the fundamental problem remains: about one third of the local baryons remain missing (e.g., \citealt{shull12}; \citealt{nicastro16}).

Future instruments, in particular, the Athena X-IFU \citep[][]{barret16,kaastra13}, are targeted at increasing the prospects to measure weak signals from hot WHIM. With that regard, in this work we aim to improve the observational status of hot WHIM absorption under the hypothesis that the warm phase ($T\sim10^5$~K, well-detected in FUV) of the WHIM is spatially co-located with the hot ($10^{6-7}$ K) WHIM (see \citealt{Nevalainen18} for more discussion). 
Other authors have previously investigated such an approach. For instance, \cite{yao09}
stacked the then available ACIS data of several AGN in order to detect
possible hot counterparts to the FUV-detected \ion{O}{VI} absorbers. \cite{Nevalainen18} used all available \emph{Chandra} ACIS and XMM-\emph{Newton} RGS data to
look for \ion{O}{VII} and \ion{O}{VIII} absorption at FUV-detected redshifts in the sight-line towards PKS2155-304. Neither study detected statistically-significant absorption, instead setting upper limits of $10^{14.5 - 15.5}$~cm$^{-2}$ of the column densities of the two ions.

In this study, we investigate the sight-line towards the quasar 3C~273 ($z\approx0.158$, \citealt{schmidt63}), also included in the \cite{yao09} sample. 
This sight-line was chosen because 
of the exceptionally good photon statistics available for high resolution X-ray spectroscopy. 
In addition, as we list in Table \ref{table:redshifts}, FUV measurements have detected \ion{O}{VI} and broad ($> 40$ km$\,$s$^{-1}$) Lyman-alpha (BLA) absorption lines at several non-zero redshifts in this sight-line (e.g., \citealt{sembach01}; \citealt{williger10}; \citealt{tilton12}; \citealt{danforth16}). 
We utilize all the currently available ACIS and RGS data in our analysis (total exposure time $>$~Ms) so that for
a single AGN we reach a similar level of sensitivity as the stacked \cite{yao09}
sample. This enables us to avoid possible problems with stacking data
from different instruments, redshifts and targets, therefore improving constraints on potentially co-located warm and hot WHIM gas.

Considering the presumed parameter space of the WHIM (i.e., $n_\mathrm{H}$, $Z$, $T$, etc.), it is not expected that photo-ionization by itself would produce high enough amounts of \ion{O}{VII} or \ion{O}{VIII} to be detectable with current X-ray instruments (e.g., \citealt{wijers18}). Since only collisional ionization seems capable of producing high enough ion column densities,
in this study we adopt the assumption that collisional ionization equilibrium (CIE) describes the physical state of hot WHIM gas.

The search for the X-ray absorbing gas is conducted at the exact redshifts where significant detections of \ion{O}{VI} absorption have been in obtained in the FUV. 
Although other authors have also used \ion{Ne}{VIII} as a tracer of hot gas (e.g., \citealt{burchett19}; see also \citealt{wijers18}), the spectral data toward 3C~273 lack the necessary wavelength coverage for such an analysis, so we do not consider it in this analysis.
After the initial search for the hot phase absorption lines at the \ion{O}{VI} absorber redshifts, each tentative line detection is critically considered by: 1) confirming the signal is consistent between all the measuring instruments, 2) checking the spectral band at the location of potential line detections against possible spectral and co-adding artifacts, and 3) comparing the freely fitted line intensity ratios to the constraints set by the CIE model. 

We complement the X-ray analysis with an optical study of the galactic filaments at the location of the absorbers, since large scale filaments are expected to harbor a major fraction of the local missing hot baryons. Finally, we compare our results to the WHIM observables predictions of the \texttt{EAGLE} \citep{schaye15} cosmological, hydrodynamical simulations.

\begin{table*}
\caption{List of FUV Absorbers Towards the 3C~273\label{table:redshifts}}
\begin{center}
\begin{tabular}{llllll}
\hline\hline
$z$ &	Line ID & $\lambda_\mathrm{rest}$ & $W$ & $b$ & $N$ \\
 &	& (\AA) & (m\AA) & (km$\,\mathrm{s}^{-1}$) & log(cm$^{-2}$)\\  
\hline     
0.0034 & \ion{O}{VI} & 1031.9  &  26.6$\pm$4.1 & $\sim$28.0 & $13.353\pm{0.077}$ \\ 
0.0073 & Ly$\alpha$  & 1215.7  &  28.0$\pm$7.0 &  45.0$\pm$14.0 & $12.740\pm{0.110}$ \\                                          
0.0076 & Ly$\alpha$  & 1215.7  &  43.0$\pm$5.0 &  60.0$\pm$10.0 & $12.940\pm{0.065}$ \\                     
0.0076 & \ion{O}{VI} & 1031.9  &  14.0$\pm$4.0 & 25.0$\pm$9.0  & $13.110\pm{0.160}$ \\                
0.0876 & Ly$\alpha$  & 1215.7  &  55.0$\pm$1.0 &  42.0$\pm$2.0 & $13.050\pm{0.020}$ \\     
0.0898 & Ly$\alpha$  & 1215.7  &  86.0$\pm$4.0 &  46.0$\pm$2.0 & $13.220\pm{0.070}$ \\   
\emph{0.09017} & \ion{O}{VI} & 1031.9  &  16.0$\pm$3.2 & 22.2$\pm$10.8 & $13.263\pm{0.110}$ \\  
\emph{0.09017} & \ion{O}{VI} & 1037.6  &  14.1$\pm$5.0 & 22.2$\pm$10.8  & $13.263\pm{0.110}$ \\                                                             
\emph{0.12005} & \ion{O}{VI} & 1031.9  &  25.3$\pm$3.0 & 10.0$\pm$3.1   & $13.437\pm{0.058}$ \\                                     
\emph{0.12005} & \ion{O}{VI} & 1037.6  &  17.8$\pm$2.9 & 10.0$\pm$3.1  & $13.437\pm{0.058}$  \\
\hline
\end{tabular}   
\end{center}
\tablefoot{FUV redshifts of significantly detected BLA's ($b>40~\mathrm{km\,s}^{-1}$) and \ion{O}{VI} absorbers in the 3C~273 sight-line \citep{tilton12}. The rest-frame line equivalent widths, $W$, line broadening parameters, $b$, and HI/\ion{O}{VI} column densities, $N$, are listed with $1\sigma$ uncertainties (if available). The emphasized redshifts were considered in this work.
}
\end{table*}

\section{Observational X-ray data set}

The absorption signatures of the hot WHIM filaments are expected to be close to or beyond the limits of detectability with the current high spectral resolution X-ray instruments. With this in mind, we analyzed all the long exposure ($\gtrsim 20$~ks) 3C~273 observations (as available on Jan 2019) from the XMM-\emph{Newton} RGS (first and second spectral orders), \emph{Chandra} LETG (HRC and ACIS first-order), and \emph{Chandra} MEG (ACIS first-order) instruments. The total RGS exposure times are $\approx 700$ ks per instrument, which is substantially longer than that of the \emph{Chandra} instruments (see Table \ref{table:data}).

Due to the shorter observation times and smaller effective area of the \emph{Chandra} instruments as compared to the RGS, the statistical weight of LETG and MEG data in the spectral modeling is much less than that of RGS data. We nevertheless found it beneficial to include the \emph{Chandra} data into  analysis, because due to the malfunctioning RGS2 CCD array \#4, there are no RGS2 first-order data in the wavelength band between $\approx 20-24$ \AA. This band covers the redshifted wavelengths of the \ion{O}{VIII}~Ly$\alpha$ line for $z\approx0.06-0.27$ and \ion{O}{VII}~He$\alpha$ for $z\lesssim0.11$ (which are the lines most likely to be detected), and because the second-order spectra of the RGS instruments have zero effective area at $\lambda \gtrsim 19$ \AA, only LETG and MEG data can provide additional information to RGS1 first-order at these wavelengths. Also, since we are examining very weak spectral signatures, examining all the available data sets is useful for identifying false positives that could occur in individual data sets due to systematic effects such as calibration errors.

\begin{table*}[t]
\caption{Observational Sample\label{table:data}}
\begin{center}
\begin{tabular}{ccc|>{\raggedleft\arraybackslash}p{8mm}c|>{\raggedleft\arraybackslash}p{8mm}cc}
\hline\hline
\multicolumn{3}{c|}{XMM-\emph{Newton} RGS} &  \multicolumn{2}{c|}{\emph{Chandra} HETG} &  \multicolumn{3}{c}{\emph{Chandra} LETG}  \\
& \multicolumn{2}{c|}{Clean time (ks)} &   & {Clean time (ks)} & & \multicolumn{2}{c}{Clean time (ks)}  \\
OID & RGS1 & RGS2 &  OID & ACIS & OID & ACIS  & HRC \\
\hline
0126700301	&	61.7 & 59.8	&	14455 & 29.6 &			460 	& 		& 39.9 \\
0126700601	&	25.9 & 25.1	&	17393 & 29.5 &			1198	& 38.2	&  \\
0126700701	&	15.9 & 15.5	&	18421 & 29.6 &			2464	& 29.5	& 29.7 \\
0126700801	&	41.5 & 40.3	&	19867 & 26.9 &			2471	& 24.9	& \\
0136550101	&	84.0 & 81.6	&	20709 & 29.6 &			3574	& 27.3	& \\
0136550801	&	50.2 & 50.2	&	2463 & 26.7 &			4431	& 26.4	& \\
0136551001	&	27.9 & 27.9	&	3456 & 24.5 &			5170	& 28.4	& \\
0137551001	&	20.5 & 19.9	&	3457 & 24.9 &&&\\	
0159960101	&	57.9 & 57.9	&	3573 & 29.7 &&&\\	
0414190101	&	60.5 & 60.5	&	4430 & 27.2 &&&\\	
0414190301	&	30.0 & 30.3	&	459 & 38.7 &&&\\	
0414190401	&	31.0 & 30.8	&	5169 & 29.7 &&&\\	
0414190501	&	37.1 & 37.1	&	8375 & 29.6 &&&\\	
0414190701	&	36.0 & 36.0	&	9703 & 29.7 &&&\\	
0414190801	&	40.7 & 40.6	&&&&	\\	
0414191001	&	38.6 & 38.6	&&&&	\\	
0414191101	&	71.1 & 71.1	&&&& 	\\
0414191301	&	64.4 & 64.4 &&&&	\\
0414191401	&	74.5 & 74.6 &&&&	\\
\hline
Co-added Spectra:    &	730 & \multicolumn{1}{c|}{723}    & & 405 & &  174  & 69 \\ 	
\hline    
\end{tabular}
\tablefoot{Total clean times of the individual and co-added spectra. The co-added spectra were used in the analysis. In the case of RGS, two co-added spectra were generated for both instruments, one for each spectral order (first- and second-orders).}
\end{center}
\end{table*}

\section{Data processing}\label{reduction}

\subsection{XMM-\emph{Newton} RGS}

The RGS data were processed with the XMM-\emph{Newton} \texttt{\small SAS} 16.0.0 software using the calibration file release \texttt{\small XMM-CCF-REL-347}. Both first- and second-order data were reduced with the \texttt{\small rgsproc} pipeline. We used somewhat more accurate processing options compared to the \texttt{\small rgsproc} defaults, in order to maximize the data quality for the examination of very weak spectral signals. Namely, we rejected the cool pixels from the data, 
and corrected for pixel-dependent energy offsets (\texttt{\small rgsproc rgsenergy} option \texttt{\small withdiagoffset=yes}). 
In addition, we decreased the aspect drift based frame grouping by a factor of two (i.e., \texttt{\small driftbinsize=0.5\arcsec}) and calculated the grating line spread functions using the full convolution space. It was confirmed that utilization of such more precise data processing options does improve the overall quality of the co-added spectra.

In addition, we created good time interval files by setting a threshold for the upper limit for the source flux, and filtered the data accordingly to remove the exposure time intervals contaminated by incidental Solar flares. We also employed the \emph{rgsproc} binning algorithm to bin the data into 20 m\AA\, wavelength bins (i.e., twice the default bin width), a scheme that oversamples the instrument energy resolution by a factor of $\approx3$. The data was already rebinned during the spectral extraction phase because it provides numerically more stable results compared to rebinning the data later from the full resolution spectra, which was desirable since we aimed to use combined data sets to analyze signals close to the limits of detectability.

After reducing the RGS data sample as described above, we produced co-added spectra separately for each RGS instrument and spectral order using the \texttt{\small rgscombine} procedure. These co-added spectra were then converted into the \texttt{\small SPEX}\footnote{http://www.sron.nl/SPEX} format with the \texttt{\small trafo} (v. 1.03) software. The \texttt{\small trafo} was used to create separate spectra for each co-added dataset, and in addition, to prepare a combined spectral file in which each of these co-added spectra were included (including the \emph{Chandra} spectra as described in the following section). Using this combined spectral file allows us to flexibly and correctly fit the data from the different instruments simultaneously with \texttt{\small SPEX} (Sects. \ref{modeling} and \ref{WHIM1}). 
This approach maximizes the data contributing to the analysis.

\subsection{\emph{Chandra} instruments}

\subsection{LETG}

The \emph{Chandra} LETG ACIS and HRC data were processed with \texttt{\small CIAO} v. 4.10 using the \texttt{\small CALDB 4.7.9} calibration files. The data reduction was conducted using the \texttt{\small chandra\_repro} script with the default parameter settings, and afterwards, the spectra of each of the instruments were co-added separately using the \texttt{\small combine\_grating\_spectra} spectral stacking tool. Both positive and negative order dispersion data were co-added when producing the ACIS and HRC spectral files for the analysis.

The statistical errors in all of the processed \emph{Chandra} LETG observations used the Gehrels approximation for Poisson confidence intervals \citep[as defined in][]{gehrels86}, which are not 
supported by the \texttt{\small SPEX} rebinning algorithm. Since the processed LETG data were 
over-sampled in the initial reduction, before converting the generated FITS data files into the \texttt{\small SPEX} format, we replaced the spectral data uncertainties with $\sqrt{N}$, where the $N$ denotes the number of counts per spectral bin. The number $N$ is in the range between $20-60$~cts/bin (HRC) and $30-150$~cts/bin (ACIS) in the studied wavelength band, meaning that the uncertainties are well approximated by $\sqrt{N}$ \citep[as is shown, e.g., in][]{bonamente2017book}.

Rebinning of the data was conducted using the \texttt{\small ftools grppha} software tool with a binning factor of two, resulting in a 30~m\AA\ bin size, or $0.6\times\Delta\lambda$ channel width (we note that \texttt{\small grppha} rebinning is carried out by means of flagging, and is not a workaround for the rebinning issue mentioned above). Like with the RGS data, we then used \texttt{\small trafo} to create the separate co-added \texttt{\small SPEX} spectral files for each LETG instrument, and to add these spectra into the combined spectral file used in the simultaneous spectral modeling.

\subsection{HETG}

The HETG observations were reduced using the \texttt{\small chandra\_repro} default processing, and the first-order data of the \emph{Chandra} HETG Medium Energy Grating (MEG) arm were extracted. The data was co-added with the \texttt{\small combine\_grating\_spectra} tool and converted into the \texttt{\small SPEX} format with \texttt{\small trafo}. Like the data from the other instruments, we added this spectrum into the combined spectral file for the simultaneous analysis.

\section{The analysis method}\label{method}

\subsection{Data qualification}\label{qualifying}

As described in Sect. \ref{reduction}, several steps were taken during data processing to minimize the risk of forming artificial absorption-like spectral features in the co-added spectra (especially in the case of the RGSs, which have the largest statistical weight in the sample). In addition, we applied two criteria throughout the analysis to further exclude sources of confusion. 
First, the closest CCD gap must be at least one HWHM away from the examined line centroid wavelength (only relevant to RGS since we co-added \emph{Chandra} positive and negative orders), and 
second, no bad pixels or columns may be present in the immediate vicinity of examined lines (or more precisely, within the FWHM of the fitted line profile).  

These criteria are important because small variations in telescope pointing (with respect to the dispersion direction) between different observations cause small shifts of the spectral wavelength scale at the detector plane. When such observations are co-added, the location of bad pixels and columns may vary in the detector wavelength space, which for a time-varying source such as 3C~273 can lead to an incorrectly defined effective area around the corresponding wavelength bins (more details of the issue can be found in \citealt{kaastra11}). As a result, spectral artifacts characteristically resembling those of absorption/emission lines may emerge in co-added spectra, or, if overlapping with an existing spectral feature, bias the measurements. 

Manual inspection of RGS2 event files revealed bad pixels coinciding with the wavelength of the $z=0.09017$ \ion{Ne}{IX} He$\alpha$ line in the RGS2 first-order data, and the data were therefore omitted from the analysis of this line. Other than that, no exclusion of data followed from these criteria for the line candidates examined in this study.

\subsection{Redshift selection}\label{redshifts}

To find the most likely locations of hot WHIM absorbers in the 3C~273 sight-line, we examined the FUV absorber data of the \cite{tilton12} FUSE+STIS survey (the sight-line has also recently been studied by \citealt{williger10}, \citealt{danforth08}, \citealt{tripp08} and \citealt{sembach01}), looking for the redshifts meeting our predetermined criteria. Namely, we were interested in the redshifts where: 
1) at least two statistically significant metal absorption lines have been  
measured in the FUV, indicating a high concentration of metals in the absorber (we note that significant detections of two lines, such as those of the \ion{O}{VI} doublet, makes it very likely that the FUV redshift is correctly determined), or 
2) a broad Lyman alpha line with $b>70$ km$\,$s$^{-1}$ has been measured, indicating that the gas temperature may be high enough to produce significant lines in the soft X-ray band (as $b=70$ km$\,$s$^{-1}$ corresponds to a $\approx3\times10^5$ K gas temperature assuming pure thermal line broadening). 

Criterion~2 was adopted because it could reveal high temperature absorbers where the \ion{O}{VI} ion fraction is too low to enable significant detection of \ion{O}{VI} lines in the FUV, although the numerical value for the $b$-limit was chosen simply to exclude number of uninteresting (low temperature) Ly$\alpha$ absorbers from the X-ray analysis.
We note, however, that without further information on the processes contributing to the Ly$\alpha$ line widths (turbulence, spectral line blending, etc), the thermal information content of the line width measurements is limited (e.g., \citealt{tepper12}).

Two redshifts met the former criterion (0.09017, 0.12005, see Table \ref{table:redshifts}). The \ion{O}{VI} line profiles at these redshifts were checked against spectral characteristics typically associated with high velocity outflows (e.g., line profile asymmetries, velocity-dependent partial coverings, time variability, line widths), to exclude the possibility that the absorption lines are related to a high velocity outflow from 3C~273. No such indications were found for any of the examined FUV lines, implying the line detections most likely relate to intergalactic absorbers. These two FUV redshifts were adopted for the X-ray follow up study.

\subsection{Modelling the 3C~273 emission and Galactic absorption}\label{modeling}

The X-ray spectra used in this analysis are combinations of multiple exposures from different epochs of a time-varying source. In such co-added spectra, the physical information carried by the continuum shape is largely averaged away. Indeed, we found that even when using limited wavelength bands, the complex shape of the quasar continuum was poorly fitted with power-law models. We therefore chose to model the emission continua of each dataset independently with a non-physical `spline' model, which is a model specifically designed for accurate modeling of X-ray continuum shapes in cases where the physics underlying the continuum shape is not sufficiently well known.
Adopting a non-physical continuum model is also useful because it can compensate for possible wider band systematics in instrument effective area models, thus improving the accuracy of the examination of weak (and narrow) absorption features in the spectra, such as those expected for the hot WHIM absorbers. 

In the analysis we used a `spline' model that consists of 17 uniformly spaced grid points between 13 and 29 \AA, thus providing 1 \AA\, intervals for the grid points; This scheme was found to enable accurate modeling of the continua, while the grid being sparse enough not to model any line features in the spectra. We allowed the `spline' $y$-parameters (i.e., the intensity) to vary throughout our analysis, so to ensure that the uncertainties in the continuum modeling were always properly propagated into the uncertainties of the model parameters of interest. 
The fitting band was fixed to $14-28$ \AA, which covers the most important transition lines of \ion{Ne}{IX}, \ion{O}{VII-VIII}, and \ion{N}{VII} for both considered FUV redshifts. 

We built a Galactic absorption model consisting of a component for the neutral disk and another one for the hot halo. For the neutral gas, we used the \texttt{\small SPEX} CIE absorption model called `hot' with the temperature fixed to $5\times10^{-4}$ keV (\texttt{\small SPEX} `neutral plasma model'), and the hydrogen column to $N_\mathrm{H}=1.77\times10^{20}$ cm$^{-2}$ \footnote{http://www.swift.ac.uk/analysis/nhtot/index.php}. Although the used hydrogen column density value has a 16~\% systematic uncertainty (see \citealt{willingale13}), we found no benefit in letting it vary during the fits, because the `spline' model is able to compensate for the possible bias in $N_\mathrm{H}$.
Instead, we found that when using the \texttt{\small SPEX} default for the line-broadening parameter ($\sigma_v=b/\sqrt{2}=100$~km$\,$s$^{-1}$), the Solar abundances highly overestimate the \ion{O}{I} lines of the component (blend at $23.5$ \AA), which is a possible indication of \ion{O}{I} line saturation. We therefore measure the \ion{O}{I} column density by letting the model oxygen abundance vary, while adopting the line-of-sight Doppler spread parameter $b=18.6$~km$\,$s$^{-1}$ for the line widths, which has been found to describe the low ionization state Galactic lines towards the 3C~273 elsewhere \citep{savage93}. We point out that even if the X-ray instruments are insensitive to line width information of narrow features ($b<$ several hundred km$\,$s$^{-1}$), the $N_\mathrm{OI}$ will be correctly constrained by the equivalent width measurement over the doublet when the value of $b_\mathrm{OI}$ is known. This fit yielded log$\,N_\mathrm{OI}(\mathrm{cm^{-2}})=16.91_{-0.09}^{+0.06}$ (corresponding $0.8$ times the Solar O abundance), a result in line with $N_\mathrm{OI}$ at other lines of sight with similar $N_\mathrm{H}$ \citep[e.g.,][]{cartledge03}. As far as we know, this is the most accurately constrained Galactic $N_\mathrm{OI}$ measurement towards 3C~273.

To model the Galactic hot halo absorption (see e.g., \citealt{fang03}), we prepared another \texttt{\small SPEX} `hot' component where the free parameters were the electron temperature and $N_\mathrm{H}$, and the elemental abundances were fixed to Solar. Fits with a hot halo component alongside the neutral gas absorber yielded $kT=0.127\pm0.007$ keV and $N_{\mathrm{H}}=4.8_{-0.5}^{+0.6}\times 10^{19}$ cm$^{-2}$ (Table \ref{table:halo}). 

The emission model with Galactic absorption, as described above, was used in all of the analysis described in the following sections. It is important to note that none of the freely varying parameters (as described above) were fixed at any point of the spectral analysis, to ensure proper error propagation in the calculation of the redshifted absorption component fit parameters. All the analyses were performed using \texttt{\small SPEX} version 3.03.00 with proto-Solar abundances of \cite{lodders09}, which is the \texttt{\small SPEX} default, and the models were fitted using Cash Statistics \citep{cash78}.

\begin{table}
\caption{Foreground Galactic Absorption Model\label{table:halo}}
\begin{tabular*}{\columnwidth}{p{0.4\columnwidth}ll}
\hline\hline
Parameter & Galactic Neutral  & Galactic Hot \\ 
\hline
$N_{\mathrm{H}}$ ($10^{19}$ cm$^{-2}$) & $17.7\,^\dagger$ & $4.8_{-0.5}^{+0.6}$\\
$kT$ (keV) & $5\times10^{-4}\,^\dagger$ & $0.127\pm0.007$  \\ 
O & $0.8\pm0.1$  &  $1^\dagger$ \\
\hline
\end{tabular*} 
\tablefoot{Fitted (and fixed) values defining the foreground Galactic absorption model used throughout the X-ray analysis. The O abundances are given in Solar units relative to H. All other elements were fixed to the Solar values. \\
$\dagger$Parameter was fixed during the minimization, see text for details.}
\end{table}

\begin{table*}[t]
\caption{X-ray Line Wavelengths\label{table:spectral_shifts}}
\begin{center}
\begin{tabular}{l|llll|l}
\hline\hline
& \multicolumn{4}{|c|}{Instrument}  \\
Line  & RGS1 & ACIS (LETG) & HRC & ACIS (HETG) &  PREDICTED \\
\hline
Galactic \ion{O}{VII} & $21.604\pm0.004$ & $21.599\pm0.005$ & $21.609\pm0.014$ & $21.611\pm0.002$ & $21.602$ \\
\hline
\ion{Ne}{IX} ($z=0.09$) & $14.67\pm0.02$ & $14.66\pm0.01$ & $14.70_{-0.10}^{+0.07}$ & $14.67_{-0.007}^{+0.01}$ & $14.66$ \\
\ion{O}{VIII} ($z=0.09$) & $20.66_{-0.02}^{+0.03}$  & $20.66\pm 0.03$ & $20.69_{-0.08}^{+0.03}$ & $20.68^1$ &  $20.68$ \\
\hline
\end{tabular}
\tablefoot{Measured line centroid wavelengths of the first-order co-added spectra. In the column PREDICTED the redshifted wavelenghts for $z=0.09017$ are listed. All values are in Angstroms. \\
$^1$ Only the best-fit value is quoted due to the low S/N ratio. }
\end{center}
\end{table*}

\section{X-ray analysis and Results}\label{WHIM1}

To verify the wavelength-scale accuracy of the co-added spectra, we first checked each of the spectra against linear shifts. This was done using the strongest spectral feature present in the 3C~273 X-ray spectrum, the Galactic \ion{O}{VII} He$\alpha$ line at $\lambda =21.602$~{\AA}. We fitted the data around this feature with a model combining a Gaussian absorption line with the `spline' continuum model, while letting the Gaussian line centroid wavelength and the line normalization parameter vary. 
We found that the centroid wavelengths of the Gaussian modeling
were consistent with the rest wavelength of \ion{O}{VII} He$\alpha$ within
the statistical uncertainties of $\lesssim 10$ m{\AA} (see Table \ref{table:spectral_shifts}). This was true for each instrument, and indicates a redshift
accuracy of $\sigma_z \approx 0.0005$ (corresponding to velocity accuracy $\Delta v\approx137$~km$\,$s$^{-1}$ at $z=0.09$, and $\approx132$~km$\,$s$^{-1}$ at $z=0.12$). We will utilize this accuracy further when
comparing the redshifts of the X-ray lines with those of the FUV lines
(Section \ref{slab}), and with those of the large scale filaments (Section \ref{filament}).

Two different models were investigated in the spectral analysis. These were the emission model with Galactic absorption (Sect. \ref{modeling}) combined with:
1) a redshifted `slab' line absorption component (Model 1, Sect. \ref{WHIM}), and 
2) a redshifted gas absorption component in collisional ionization equilibrium with solar relative abundances, (Model 2, Sect. \ref{cie}). 

In addition to these models, we used a model combining Gaussian lines and the Galactic absorption model whenever determining absorption line centroid wavelengths in the spectra: This approach was used to avoid shifting of the `spline' emission component energy grid, which would take place if letting the model redshift to vary during the fits.\\

The initial search for redshifted absorption lines was conducted by fitting Model 1 independently to the co-added spectra of different instruments and spectral orders, and then by fitting the co-added spectra simultaneously with the same model. Considering all these measurement results together, we searched for inconsistencies between the results given by different instruments, and if such were found, the hypothesis of an astrophysical line origin was rejected. 

Model 1 was also used in examining the effects of systematics on the measurement results of the putative redshifted lines. Namely, we re-fitted the RGS1 first-order data, which has the largest statistical weight in our sample, after adding a $2\%$ Poissonian noise component to the data (corresponding to the systematic calibration uncertainty level of RGS first-order data, see \citealt{vries15}).
We found this noise component to have negligible effect on the fit parameters of interest, and therefore omitted its use in the further analysis.

After mapping the redshifted line candidates, the main analysis of this work was conducted using the simultaneous fitting method with Model~2, ensuring that all the information in each of the co-added spectra was taken into account in the analysis. The simultaneous fitting of data from different instruments and spectral orders was conducted with the appropriate \texttt{\small SPEX} tools\footnote{http://var.sron.nl/\texttt{\small SPEX}-doc/manualv3.03.00.pdf}. Technically, in these fits, all the absorption components were coupled (i.e., both Galactic and redshifted components) between the models of the different co-added spectra, while the `spline' continuum emission models were uncoupled. With this method, the continua of all the separate datasets, observed at different epochs with different instruments, were fitted at a level which enabled the investigation of weak absorption components at high accuracy (we get C-statistics/d.o.f.$<1.07$ in the used $14-28$~\AA\, fitting band).

\subsection{The redshifted absorption line model (Model 1)}\label{WHIM}

To search for X-ray absorption lines in the data, we added a redshifted \texttt{\small SPEX} `slab' component to the absorbed emission model described in Sect. \ref{modeling}. 
The `slab' component models the transmission of photons through an optically thin layer of gas, and enables measurement of column densities of different ion species while setting no constraints on the gas ionization balance. The `slab' component does, however, include the relative intensities of the spectrum of transition lines produced by the examined ion, thus providing more accurate results as compared to modeling the data with a single line profile. The \texttt{\small SPEX} `slab' absorption model fits the lines with the Voigt profile, in which the Gaussian component has an adjustable line broadening to account for the thermal and non-thermal velocity distributions of the absorbing gas. Because we did not have information on either of these components, and since the used instruments' energy resolution is insufficient to model the line shapes, we fixed the `slab' line width parameter to the \texttt{\small SPEX} default value of $100$ km$\,$s$^{-1}$ at all times (corresponding $T\sim10^7$~K thermal broadening for oxygen lines).

The spectral analysis was conducted by fitting the ion column densities of the `slab' model's \ion{O}{VII-VIII}, \ion{Ne}{IX}, \ion{Fe}{XVII}, and \ion{N}{VII} ions one by one, while fixing all the other `slab' ion columns to zero during each fit. Using this approach, we investigated both of the FUV guided redshifts (0.09017, 0.12005) by fixing the redshift to those values. We note that the major benefit of using the `slab' model to search for possible absorption lines is that it can be used to obtain the ion columns without any knowledge of the gas temperature. Therefore the analysis with the `slab' model set up useful limits for the examined ion column densities, whose credibility can be tested against the predictions of physically justified models, such as CIE.

\begin{figure}
\begin{center}
\includegraphics[width=\linewidth]{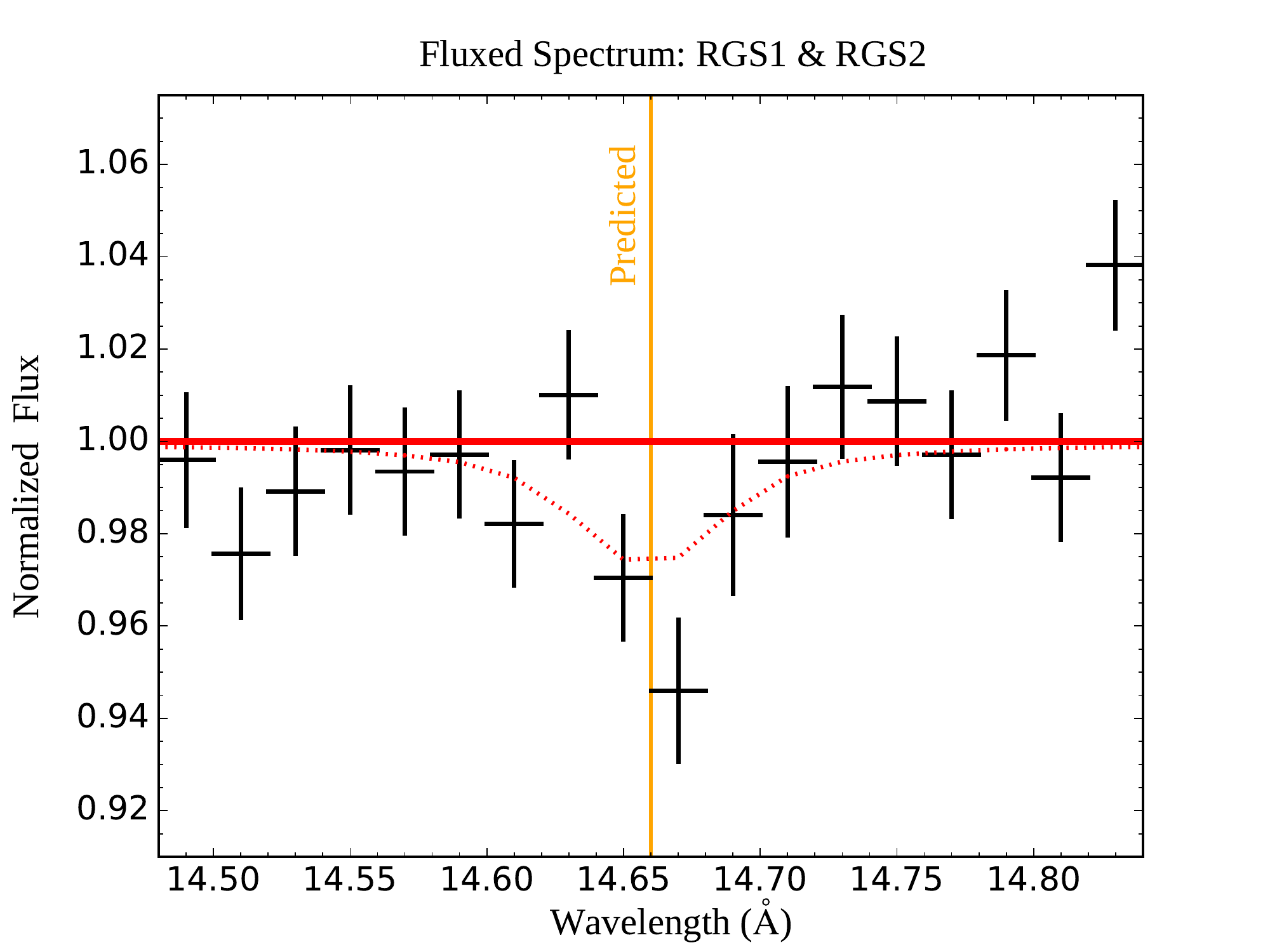}
\end{center}
\caption{Continuum normalized, co-added fluxed spectrum of RGS1 and RGS2 data at the wavelength around the putative redshift 0.09017 \ion{Ne}{IX} He$\alpha$ absorption line.
The FUV predicted line centroid wavelength is shown with the orange line, while the dotted red curve shows the expected absorption signal based on the simultanous fitting of XMM-\emph{Newton} and \emph{Chandra} data with the redshifted $slab$ model (Sect. \ref{slab}). }
\label{fig:fluxed}
\end{figure}

\subsubsection{Absorption line analysis results: z=0.09017}\label{slab}

We began the line analysis at the \ion{O}{VI} absorber redshift $z=0.09017$. Because the hot WHIM absorption lines are expected to have equivalent widths of only $\lesssim10$ m\AA\, \citep[e.g.,][]{oppenheimer16,wijers18}, the visual prominence of possible WHIM lines in the co-added spectra is expected to be weak. For this reason, any existing line features could easily be overlooked and interpreted as noise. However, if the absorption features do have an astrophysical origin, then one can improve their visual appearance by utilizing the procedure introduced in \cite{kaastra11}. Following this method, the SAS program \texttt{\small rgsfluxer} was used to generate fluxed RGS1 and RGS2 spectra for each spectral order, which were then combined using the \texttt{\small SPEX} tool \texttt{\small rgsfluxcombine} to generate a single fluxed spectrum. 
We note that producing a fluxed spectrum also works as an independent sanity check, because the tool is less prone to produce stacking artifacts than \texttt{\small rgscombine}. We found that the resulting spectrum with increased photon statistics (at the wavelength bands where available) indeed shows a strengthened absorption feature at the wavelength $\approx14.66$~\AA, which corresponds to the redshifted wavelength of the \ion{Ne}{IX} He$\alpha$ line at $z=0.09017$ (Fig. \ref{fig:fluxed}).

Quantitative analysis of \ion{Ne}{IX} by simultaneous fitting of Model 1 to the co-added XMM-\emph{Newton} and \emph{Chandra} data yielded $W=2.7\pm0.9$ m{\AA} for the rest-frame equivalent width of the \ion{Ne}{IX} He$\alpha$ line, corresponding to an ion column density log$\,N_\mathrm{\ion{Ne}{IX}}(\mathrm{cm^{-2}})=15.4_{-0.2}^{+0.1}$ for \ion{Ne}{IX}. In Fig. \ref{fig:fluxed} we plot this predicted absorption signal for the \ion{Ne}{IX} He$\alpha$ line on top of the fluxed spectrum for comparison.
In addition, we found an absorption line feature at the wavelength corresponding the $z=0.09017$ \ion{O}{VIII} Ly$\alpha$ line, for which we measured a rest-frame $W=4.3\pm1.6$ m{\AA} corresponding log$\,N_\mathrm{OVIII}(\mathrm{cm^{-2}})=15.5\pm0.2$. In case of \ion{O}{VIII} Ly$\alpha$, the application of the fluxed spectrum method was not useful, because the corresponding wavelength band is only observable in the RGS first-order spectrum and accordingly no improvement of photon statistics was available. We point out that the \ion{O}{VII} He$\alpha$ line, which is typically expected to be the strongest X-ray WHIM line, would blend with the (strong) Galactic \ion{O}{I} line ($\lambda_{z=0}^{\mathrm{O\,I}}\approx23.51$ \AA, $\lambda_{z=0.09017}^{\mathrm{O\,VII}}\approx23.55$ \AA), thus preventing  the analysis of this line. Accordingly, we are only able to report here the weak upper limits for \ion{O}{VII} at $z=0.09017$ based on the fits to the non-blended transition lines, including the \ion{O}{VII} He$\beta$ and others. 

\begin{figure*}[t]
\includegraphics[width=\textwidth]{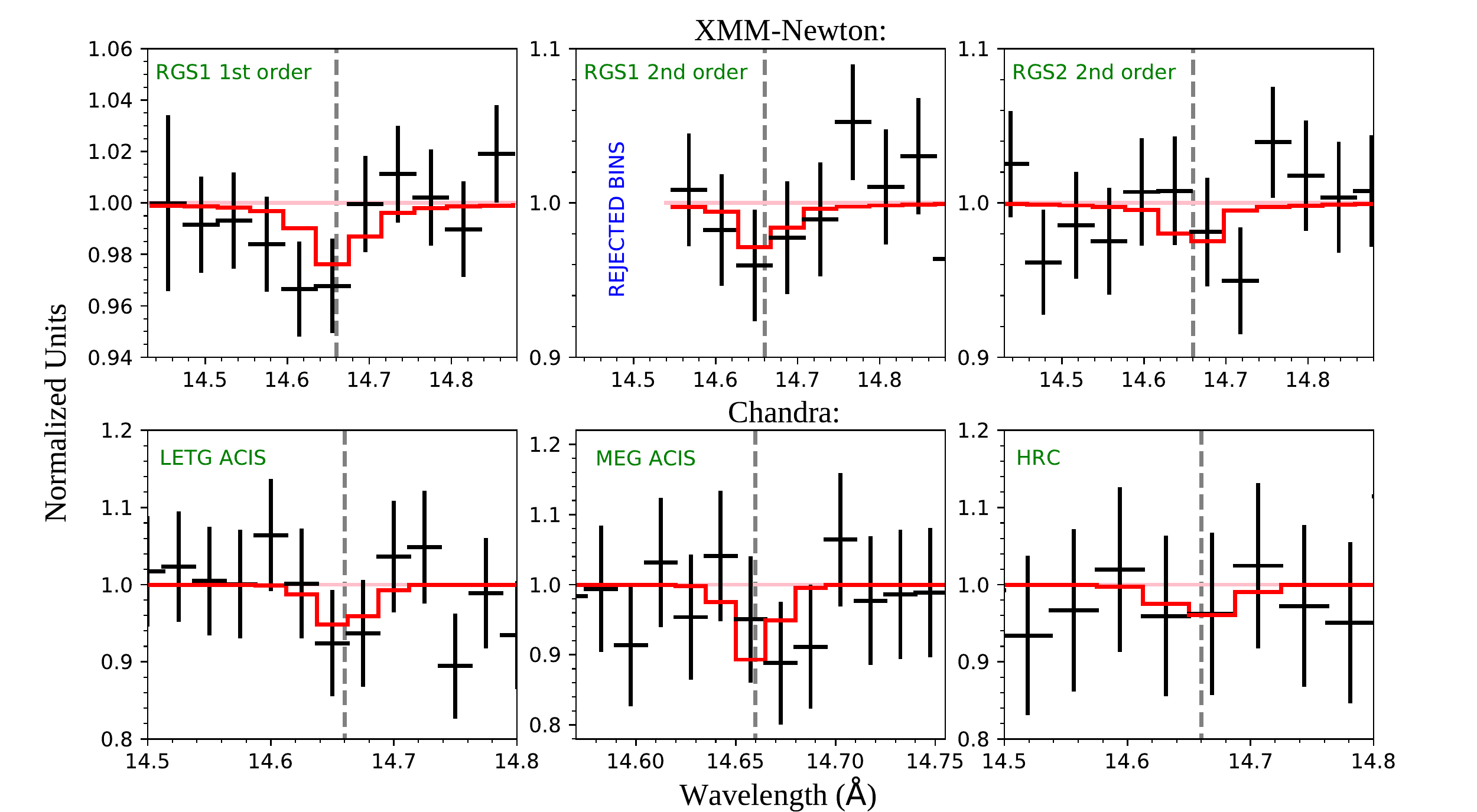}
\caption{Normalized data (crosses) and the best-fit model (red line) of the $z=0.09017$ \ion{Ne}{IX} line (the FUV predicted line centroid wavelength is marked with a dashed line) for all the instruments and spectral orders employed. The normalization is done by dividing the spectra by the best-fit continuum model. The best-fit model was simultaneously fitted to RGS1 (first and second dispersion orders), RGS2 (second-order), LETG ACIS (first-order), MEG ACIS (first-order) and HRC data, while the RGS2 first-order data was omitted due to a bad column near the centroid wavelength of the line. The missing bins in RGS1 second-order were rejected by the reduction software due to bad response. The spectral data was rebinned for this illustration to improve the S/N -ratio. }
\label{fig:Neix_all}
\end{figure*}

\begin{figure*}[t]
\includegraphics[width=\textwidth]{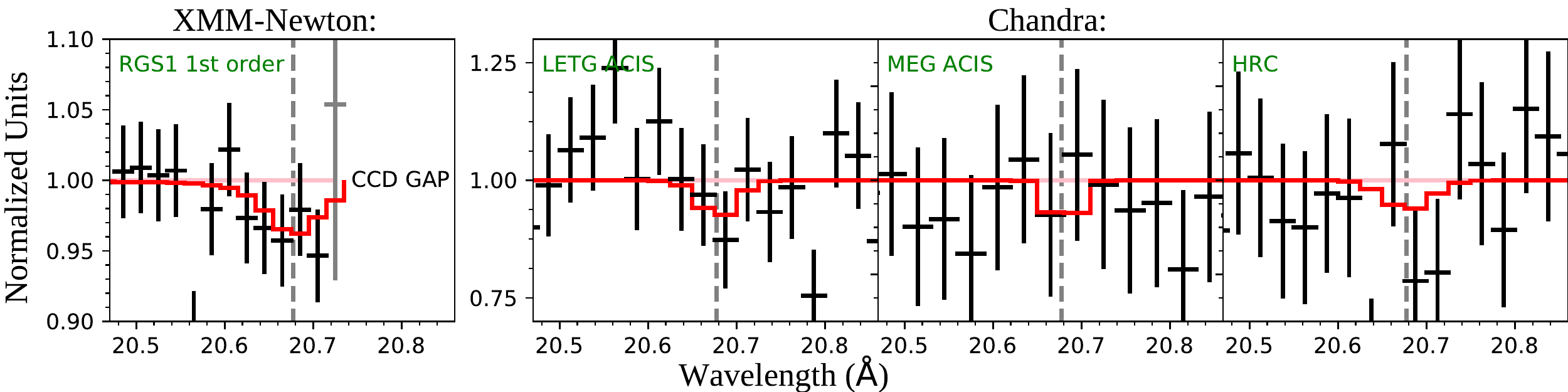}
\caption{As Fig. \ref{fig:Neix_all} but for \ion{O}{VIII}. The grey bin near the border of the RGS1 CCD gap was not used in the minimization, but demonstrates a type of spectral artifact that can be produced when observations from different epochs and slightly different pointings with respect to the dispersion direction are co-added.}
\label{fig:Oviii_all}
\end{figure*}

To double-check the measurement results obtained from simultaneous fitting with Model 1 (see the best-fit model on top of each of the individual co-added spectra in Figs.~\ref{fig:Neix_all} and~\ref{fig:Oviii_all}), we analyzed each of the co-added spectra separately with the same model. Similarly to the simultaneous modeling, the separate fits yielded ion column densities of the order of $10^{15}$ cm$^{-2}$ for both ions (\ion{Ne}{IX}, \ion{O}{VIII}) in each co-added dataset, and no inconsistencies were found between different instruments. We present the complete line analysis results, including both the simultaneous and the individual fits to the co-added spectra, in the uppermost panels of Table~\ref{table:news}.  

The co-added spectra were also studied independently by adopting a model combining the absorbed emission model (Sect. \ref{modeling}) with Gaussian absorption lines, to confirm that the spectral features suspected as redshifted \ion{Ne}{IX} and \ion{O}{VIII} lines were indeed centered at the wavelengths predicted by the FUV redshift. The Gaussian line model was adopted for this analysis because of the technical limitations of `slab' modeling in this context, as mentioned in Sect. \ref{WHIM1}.
In the fits we set the line parameters according to the results discussed above, and re-fitted the data while allowing the Gaussian line centroid wavelengths and line normalization parameters to vary.
We found that within the error margins, the Gaussian lines yielded centroid wavelengths consistent with those predicted by the FUV redshift, and that this was true for both examined lines for each available instrument (see Table \ref{table:spectral_shifts}). Since the wavelength-scale of the X-ray data was found to be accurate to $\sigma_z \approx 0.0005$, or $\Delta v_\mathrm{X-ray} \approx137$~km$\,$s$^{-1}$ (Section \ref{WHIM1}), and as the FUV redshifts are only expected to have a $z$ uncertainty of $\Delta z_\mathrm{FUV} \sim 3\times10^{-5}$ ($\Delta v_\mathrm{FUV} \approx8$~km$\,$s$^{-1}$), the hypothesis of spatial co-location of the FUV and X-ray absorbers was found to be supported by all the observational data.

Finally, since it is possible that spectral co-addition produces small artifacts which can be erroneously interpreted as astrophysical lines, and because the two tentative X-ray lines only have equivalent widths of few m{\AA}, we chose to investigate this issue further. Here we used the RGS1 first-order data, which contains the majority of the statistical weight in the simultaneous fits and is co-added from 19 different observations. We refitted the RGS1 first-order data using the original, non-stacked spectra and applied the same method used in the simultaneous fits to the spectra of different instruments and spectral orders (i.e., coupled absorption - uncoupled emission). The results obtained for both absorption lines were practically identical to the fits to the co-added RGS1 data (Table \ref{table:news2}), thus confirming that the examined spectral features are not co-adding artifacts. 

To conclude the results of the line analysis, Model 1 yielded $N_\mathrm{ion}\sim 10^{15}$ cm$^{-2}$ ion column densities for \ion{O}{VIII} and \ion{Ne}{IX}, which are in the range expected for hot WHIM absorbers. 
We found that all the instruments agree with these results, and that the features appear in the spectra at the wavelengths predicted by the FUV absorber redshift. 
The \ion{O}{VIII} and \ion{Ne}{IX} ion fractions peak at coinciding temperature ranges (i.e., $T\approx10^{6-6.5}$~K), and hence the two ions are likely indicators of a common gas phase for which the `slab' modeling yields a $3.9\sigma$ (quadratically summed) detection level.
We therefore adopted the \ion{Ne}{IX} and \ion{O}{VIII} lines as hot WHIM line candidates at the \ion{O}{VI} absorber redshift $z=0.09017$. The obtained measurement results regarding  these line candidates are next compared to the constraints from the physical WHIM absorption model (Model 2) in Sect. \ref{cie}.

\begin{figure}
\begin{center}
\includegraphics[width=\linewidth]{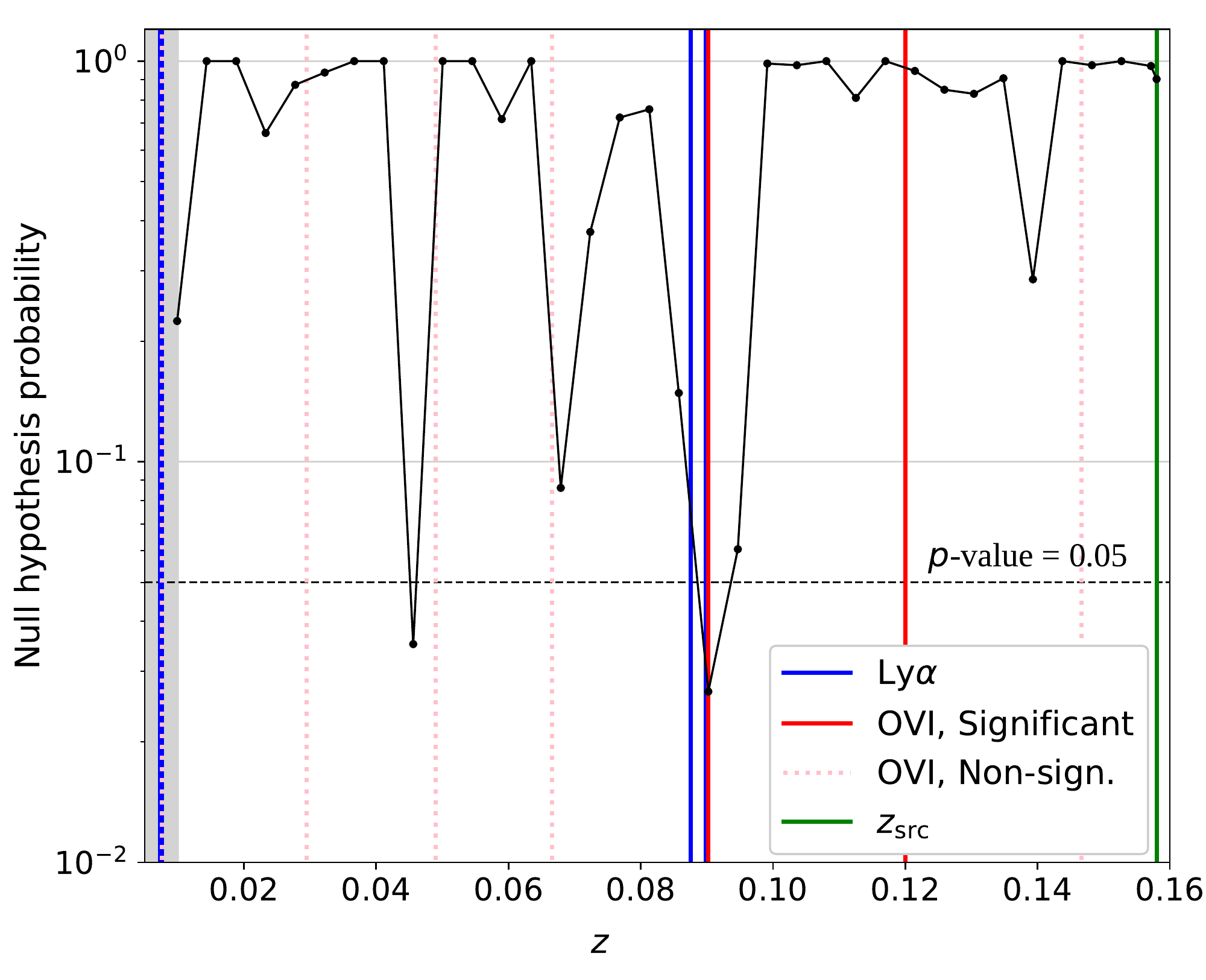}
\end{center}
\caption{Statistical probability of the Null hypothesis validity over the redshift range  $0.005-0.158$. The Null hypothesis is the emission model with Galactic absorption, which is compared to the WHIM hypothesis, i.e., a model including a redshifted hot ($T=10^{6-7}$~K) CIE absorption component. The dashed black line marks the commonly adopted Null hypothesis rejection limit, $p=0.05$. 
The FUV redshifts listed in Table~\ref{table:redshifts} are shown together with the non-significant \ion{O}{VI} doublet detections (from Table 4 in \citealt{tilton12}). The shaded color marks the confusion region of the redshifted absorption component with the Galactic hot halo.
 }
\label{fig:cvsz}
\end{figure}

\subsubsection{Absorption line analysis results: z=0.12005}\label{12}

The `slab' analysis for the FUV redshift $z=0.12005$ revealed no hot absorption line candidates at the predicted centroid wavelengths of any of the lines of interest (as listed in Sect. \ref{WHIM}), making it uninteresting for the further spectral analysis.
The X-ray non-detection is concordant with the earlier results of \cite{tripp08}, where it was found that the \ion{O}{VI}, HI Ly$\alpha$ line widths yield $T<10^{4.7}$~K (3$\sigma$ upper limit) for the FUV detected gas at the 3C~273 $z=0.12$ absorber. These X-ray and FUV results may indicate that this absorber only has a warm component.

Furthermore, we note that the non-detection of hot phase ions at $z=0.12$ means, for instance, that the 1$\sigma$ limit on the \ion{O}{VI} absorber associated $N_\mathrm{OVIII}$ 
reduces from log$\,N_\mathrm{OVIII}(\mathrm{cm^{-2}})=15.7$ (as measured at $z=0.09$) to $15.1$ when both of the \ion{O}{VI} absorber redshifts are considered simultaneously. This limit agrees well with the earlier results of \cite{yao09}, where they measured a log$\,N_\mathrm{OVIII}(\mathrm{cm^{-2}})=15.63$ $95\,\%$ upper limit on \ion{O}{VIII} columns associated with the strong \ion{O}{VI} WHIM absorbers. This upper limit comes from an analysis of $z_\mathrm{OVI}$ blueshifted, stacked X-ray spectra from 6 different sight-lines (refer to \citealt{yao09} for details). We note however, that while the $z_\mathrm{OVI}$ based stacking of the X-ray data increases the spectral S/N -ratio, the sensitivity to measure the hot phase ion signals may be compromised in this method. This is because X-ray band detectable ion columns are produced in more limited WHIM parameter space region than FUV detectable \ion{O}{VI} columns, and hence frequent non-detections of hot ion species are expected, which can effectively hide the X-ray signals in the $z_\mathrm{OVI}$ shifted, stacked spectra. 
We will investigate the co-occurrence rate of detectable \ion{O}{VI} and \ion{O}{VIII} ion columns in the WHIM absorbers in more detail in Sect. \ref{comparison}.

\subsection{Collisional ionization equilibrium modeling (Model 2)}\label{cie}

The analysis with the `slab' absorption model (Sects. \ref{slab} and \ref{12}) yielded two absorption line candidates at the wavelengths matching those of \ion{Ne}{IX} and \ion{O}{VIII} lines at the FUV absorber redshift $z=0.09017$ (Table \ref{table:spectral_shifts}). 
The ionization temperatures of \ion{Ne}{IX} and \ion{O}{VIII} are similar, 
and indicate $\mathrm{log}(T(K)) > 6$ for the absorbing gas. Since at these temperatures  photo-ionization is expected to be less important and collisional ionization equilibrium (CIE) is likely, we tested this hypothesis and modeled the data by adding a redshifted CIE absorption component with solar relative abundances to the emission model with Galactic absorption (hereafter CIE WHIM). The CIE WHIM modeling was performed using the simultaneous fitting of the co-added spectra, except that of the RGS2 first-order, which was omitted from the analysis due to the absence of reliable data at the wavelengths of either of the two putative absorption lines. 

To ensure that the CIE WHIM model robustly detects the weak signals of interest, we utilized the model in a blind search test over the redshift range $z=0.005-0.1578$.
 In this test, the lower $z$ limit was chosen large enough to avoid considerable blending with the absorption lines of the hot Galactic halo, while the upper $z$ limit corresponds to the nominal redshift of 3C~273. The redshift range was examined in steps so that each $z$-increment corresponded to a 60 m\AA\, wavelength shift of the \ion{Ne}{ix} line centroid in the spectra. At each step the co-added spectra were fitted with the CIE WHIM model, and the fit C-statistics compared to that obtained without the redshifted component included (Null hypothesis). The improvement of the fit statistics due to the addition of the redshifted absorption component were converted into a Null hypothesis probability using a modified F-test (modified, since we used C-statistics instead of $\chi^2$  as a figure of merit). The results of this test are illustrated in Fig. \ref{fig:cvsz}. As might be expected based on the results of the `slab' analysis, we found that the lowest $p$-value is indeed obtained at $z\approx0.09$ ($p\approx0.027$), while at $z\approx0.12$ the Null hypothesis is valid. 
 These findings support the applicability of the CIE WHIM modeling in the context of this study. 
 
In addition to $z=0.09$, the $p$-value falls below the often adopted Null hypothesis rejection limit, $p=0.05$, at $z\approx0.045$ ($p\approx0.035$). This redshift does not fulfill our predetermined FUV $z$ -criteria (Sect. \ref{redshifts}), but we will make a few remarks on this minimum here. The best-fit CIE model is described by $kT\approx0.10$~keV ($\approx1.1\times10^6$~K) and $N_\mathrm{H}\approx4.6\times10^{18}$~cm$^{-2}$, which together define a model producing one detectable absorption line in the examined spectral band (\ion{O}{VII} He$\alpha$, EW$\approx5$~m\AA). Inspection of the spectra reveals that this model is minimized to a line-like feature present in the RGS1 first-order data at $\lambda=22.55$~\AA, while the other instruments' data lack the sensitivity to confirm the existence of this feature.  
We note that the best-fit model does not predict any detectable signal in the FUV band (e.g., log$\,N_\mathrm{OVI}(\mathrm{cm}^{-2})<13$) and that there are no FUV detections at the corresponding redshift. In the absence of such supporting information, and since the best-fit CIE model is effectively a single line model, it remains uncertain whether the fit reveals CIE state \ion{O}{VII} absorption at $z\approx0.045$, or some other (non-CIE) ionic line at a different redshift. 
Lacking additional observational evidence to support the physical origin of this feature, or to verify that the suspected redshift is correct, we will not consider the possible $z\approx0.045$ signal further in this work.

\begin{figure}
\begin{center}
\includegraphics[width=\linewidth]{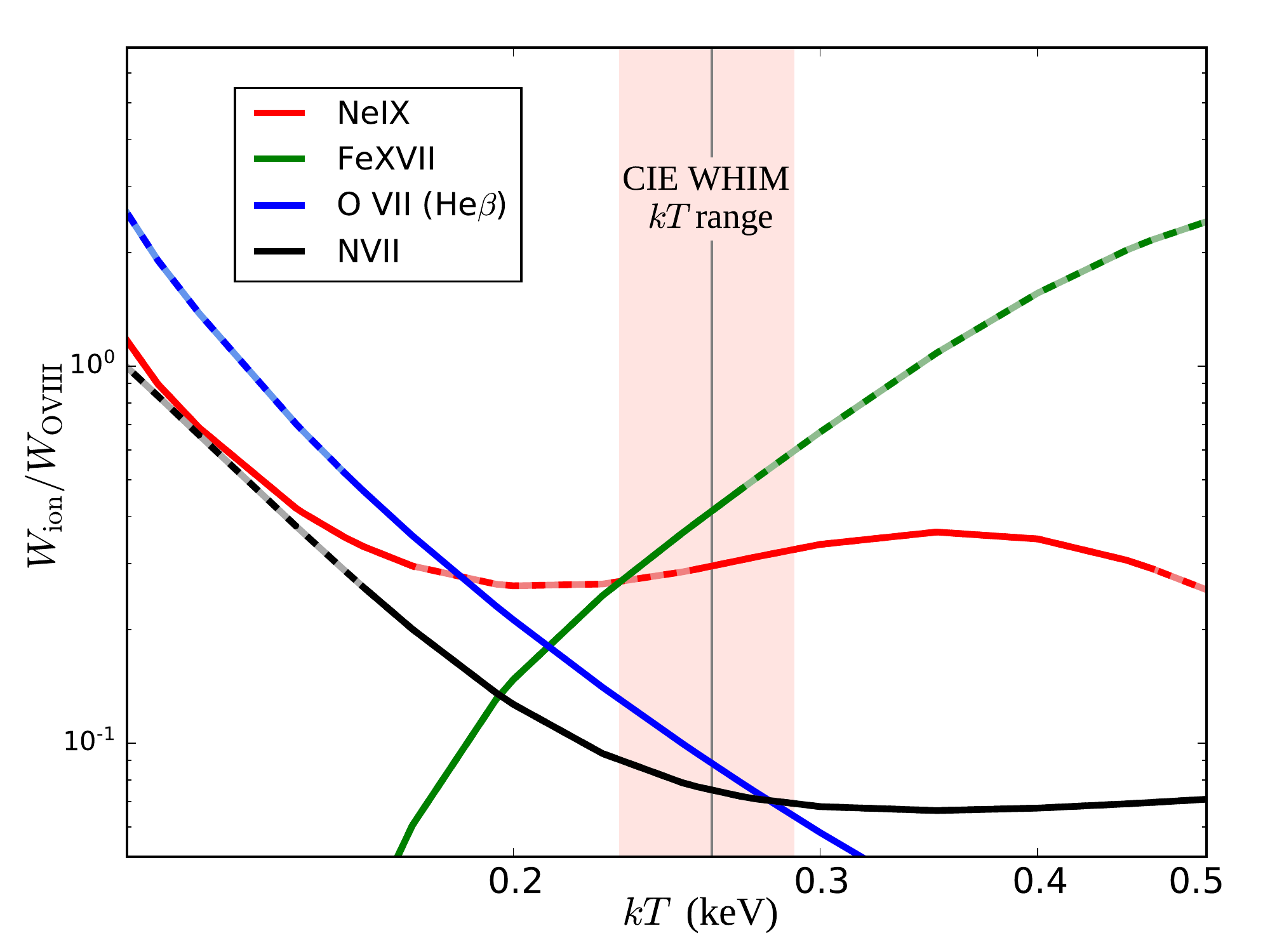}
\end{center}
\caption{Origin of temperature constraints in CIE WHIM modeling. The figure illustrates CIE model temperature dependencies of various ionic line strengths relative to that of the \ion{O}{VIII} Lyman $\alpha$ line. 
The dashed sections mark the regions where the $W_\mathrm{ion}/W_\mathrm{OVIII}$ ratio is
inconsistent with the $1\sigma$ limits of $slab$ modeling. The ions shown in the figure produce the most important absorption lines in the spectra in the examined wavelength band of 14-28 \AA\, and thus constrain the CIE temperature, marked with the shaded region. The data of the strongest available ionic lines were used to generate this figure.
}
\label{fig:constraints}
\end{figure}

\subsubsection{CIE absorber analysis results: z=0.09017}\label{cie_results}

The analysis results of the CIE WHIM modeling at the FUV redshift $z=0.09017$ are listed in the bottom panel of Table \ref{table:news}. The CIE model yielded a gas temperature $kT=0.26\pm0.03$ keV ($\mathrm{log}T\mathrm{(K)}=6.48\pm0.05$) and hydrogen column density $N_\mathrm{H}=1.3_{-0.5}^{+0.6}\times10^{19}$~cm$^{-2}$ (when $Z=Z_\sun$), corresponding ion column densities log$N_{\mathrm{NeIX}}\,(\mathrm{cm}^{-2}) =14.9\pm0.2$ and log$N_{\mathrm{OVIII}} \,( \mathrm{cm}^{-2}) =15.4\pm0.2$. We note that no other lines with comparable equivalent widths were predicted in the fitting band, nor in the wavelength range accessible to any of the used instruments (i.e., $\lambda\approx3-175$ {\AA}). 
We find that while the `slab' and CIE modeling yield matching ion columns for \ion{O}{VIII} (log$N_{\mathrm{OVIII}}^{slab}(\mathrm{cm^{-2}})=15.5\pm0.2$, log$N_{\mathrm{OVIII}}^\mathrm{CIE}(\mathrm{cm^{-2}})=15.4\pm{0.2}$),
 the CIE-predicted \ion{Ne}{IX} column is lower than obtained with `slab' modeling (log$N_{\mathrm{NeIX}}^{slab}=15.4_\mathrm{-0.2}^{+0.1}$, log$N_{\mathrm{NeIX}}^\mathrm{CIE}=14.9\pm{0.2}$).
 Such discrepancy could be explained, for instance, by \ion{Ne}{IX} excess within the particular integration path through the absorber, or as an indication of small deviations from the ionization equilibrium at the absorbing gas.
 We will consider this discrepancy further in Sect.~\ref{comparison}.
 
 We also find that the CIE WHIM model predicted \ion{O}{VI} column density, log$N_{\mathrm{OVI}}^\mathrm{CIE}(\mathrm{cm^{-2}})=12.2\pm{0.2}$, is an order of a magnitude lower than indicated by the FUV data, log$N_{\mathrm{OVI}}^\mathrm{FUV}(\mathrm{cm^{-2}})=13.263\pm{0.110}$. The measured \ion{O}{VI} absorption cannot therefore be explained by hot, collisionally ionized gas. The same conclusion can also be drawn from the FUV measured \ion{O}{VI} line broadening (Table \ref{table:redshifts}), which is smaller than thermal broadening at CIE WHIM model temperatures. We will examine these discrepancies in detail in Sects. \ref{phases} and \ref{comparison}. 

\begin{table*}[t]
\caption{Reshifted Absorber at $z=0.09017$\label{table:news}}
\begin{tabular*}{\linewidth}{lccccccc}
\hline\hline
Parameter 				& SIMULT. FIT	& RGS1 1st 				& RGS1 2nd &  RGS2 2nd & LETG ACIS & LETG HRC  & MEG ACIS  \\
\hline
& & & & \textbf{\ion{Ne}{IX}} & & & \\
log$N_\mathrm{NeIX}(\mathrm{cm^{-2}}$)  & $15.4_{-0.2}^{+0.1}$ & $15.5_{-0.4}^{+0.2}$ 	& $<15.8$ & $<15.6$ & $15.6_{-0.5}^{+0.2}$ 	& $<15.9$ 	& $15.3_{-0.6}^{+0.25}$		 \\
$W_\mathrm{He\alpha}$ (m\AA) 				& $2.7\pm 0.9$	& $3.4\pm 1.8$		& $2.6_{-2.6}^{+2.9}$ 	& $0.9_{-0.9}^{+2.8}$ 	& $4.1_{-2.6}^{+2.5}$ 			& $2.0_{-2.0}^{+5.0}$ 		& $2.4_{-1.7}^{+1.6}$			 \\
$\sigma$ 				& 2.9	& 1.8 				&  	-	  &  -	& 1.4 																&  		-					& 1.4						 \\
c-stat/dof 			& 3691/3451		& 642/557 			& 266/220 		 & 231/197 			& 548/539 									& 546/539 					& 1434/1379					 \\
\hline
& & & & \textbf{\ion{O}{VIII}} & & & \\
log$N_\mathrm{OVIII}(\mathrm{cm^{-2}}$) & $15.5\pm0.2$	& $15.5_{-0.3}^{+0.2}$ 	& n/a & n/a 	& $15.6_{-0.7}^{+0.3}$  									& $15.9_{-0.5}^{+0.3}$ 		& $<15.6$		 \\
$W_\mathrm{Ly\alpha}$ (m\AA)	 			& $4.3\pm 1.6$	& $4.3_{-2.1}^{+2.2}$ 			& n/a  & n/a 	& $4.9_{-3.8}^{+3.9}$ 										& $9.6_{-6.6}^{+6.9}$ 		& $1.1_{-1.1}^{+4.1}$			 \\
$\sigma$ 				& 2.6	& 2.1 					& n/a  & n/a 	& 1.3  														& 1.5 						&-							 \\
c-stat/dof 			& 3178/3026		& 641/557 				& n/a  & n/a 	& 549/539 													& 544/539 					& 1436/1379					 \\
\hline
& & & & \textbf{\ion{O}{VII}} & & & \\
log$N_\mathrm{OVII}(\mathrm{cm^{-2}}$) & $<15.5\,^*$ & & &  &  & &   \\
$W_\mathrm{He\alpha}$ (m\AA)  & $<8.6\,^*$  & &&  & &  &  \\
c-stat/dof  & 3146/2982 &  & & & & &   \\
\hline
& & & & \textbf{CIE} & & & \\
$kT$ (keV) & $0.26\pm0.03$ &&&&&&  \\
$N_\mathrm{H}$ ($Z_\sun/Z\times 10^{19}$ cm$^{-2}$)  & $1.3^{+0.6}_{-0.5}$ &&&&&&   \\
log$N_{\mathrm{Ne\,IX}}(\mathrm{cm^{-2}})$ & $14.9\pm0.2$ &&&&&& \\
log$N_{\mathrm{O\,VIII}}(\mathrm{cm^{-2}})$ & $15.4\pm 0.2$ &&&&&&  \\
log$N_{\mathrm{O\,VII}}(\mathrm{cm^{-2}})$ & $14.8\pm 0.2$ &&&&&&  \\

log$N_{\mathrm{O\,VI}}(\mathrm{cm^{-2}})$ & $12.2\pm0.2$ &&&&&& \\
c-stat/dof & 5156/4850 &&&&&& \\
\hline
\end{tabular*}
\tablefoot{Spectral modeling of the $z=0.09017$ X-ray absorber. The results of the `slab' modeling of \ion{Ne}{IX} ($\lambda_\mathrm{0}=13.447$ \AA), \ion{O}{VIII} (blend at $\lambda_\mathrm{0}\approx18.97$ \AA) and \ion{O}{VII} ($\lambda_\mathrm{0}=21.602$ \AA) are shown in the three uppermost panels; The rest-frame line equivalent widths of the most important absorption lines are denoted by $W$, while the $\sigma$ refers to the statistical detection significance level of the ion as yielded by the \emph{slab} modeling. The first column reports the results obtained from a simultaneous fit to the individual, co-added data sets, whereas the last 6 columns list the results when these co-added data sets are modeled separately. 1$\sigma$ upper limits for `slab' column densities are quoted if the line was not detected in individual spectra.
 The $z=0.09017$ CIE modeling results are listed in the bottom panel.  \\ 
The n/a refers to zero effective area around the wavelengths of interest. The RGS2 first-order data was not included in the fits as it misses the data around the wavelengths of all the examined line features. All the quoted uncertainties correspond to $1\sigma$ confidence levels.\\
$^*$Due to the spectral line blending with Galactic \ion{O}{I} lines, the \ion{O}{VII} upper limits were derived excluding the band of the \ion{O}{VII} He$\alpha$ line.}
\end{table*}

\begin{table}
\caption{Fits to Stacked and Non-stacked Data \label{table:news2}}
\begin{tabular*}{\columnwidth}{p{0.4\columnwidth}p{0.3\columnwidth}c}
\hline\hline
Parameter & Individual & Co-added \\
\hline
$W$, \ion{Ne}{IX} (m\AA) & $3.3\pm1.8$ & $3.4\pm 1.8$ \\
log$N_{\mathrm{Ne\,IX}}(\mathrm{cm^{-2}})$ & $15.51_{-0.37}^{+0.22}$ & $15.51_{-0.37}^{+0.22}$\\
$\sigma$ & 1.8 & 1.8 \\
c-stat/dof & 11270/10593 & 642/557 \\
\hline
$W$, \ion{O}{VIII} (m\AA) & $4.1_{-2.1}^{+2.2}$ & $4.3_{-2.1}^{+2.2}$ \\
log$N_{\mathrm{O\,VIII}}(\mathrm{cm^{-2}})$ & $15.51_{-0.32}^{+0.20}$ &  $15.54_{-0.30}^{+0.19}$ \\
$\sigma$ & 1.9 & 2.1 \\
c-stat/dof & 11269/10593 & 641/557 \\
\hline
\end{tabular*}
\tablefoot{Comparison between the results of simultaneous fits to the 19 individual RGS1 first-order spectra (column "Individual"), and the results obtained when fitting the co-added spectrum of the same data-set (column "Co-added"). The $W$ denotes the rest-frame equivalent widths of the major transition line of the ion, while the $\sigma$ marks the nominal detection significance levels for the fitted lines.}
\end{table}

The temperature of the CIE WHIM model is remarkably tightly constrained. This can be understood through the constraints set by both the detections and non-detections of the most important absorption lines in the temperature range in question (see Fig. \ref{fig:constraints}). 
Namely, the line equivalent width ($W$) ratio between the \ion{Fe}{XVII} resonance line and \ion{O}{VIII} Ly$\alpha$ is very sensitive to the temperature, and thus the non-detection of \ion{Fe}{XVII} sets a strict upper limit on the temperature. At low temperature end, non-detections of \ion{O}{VII} and \ion{N}{VII} lines set constraints for the model temperature. 

\begin{figure}
\begin{center}
\includegraphics[width=\linewidth]{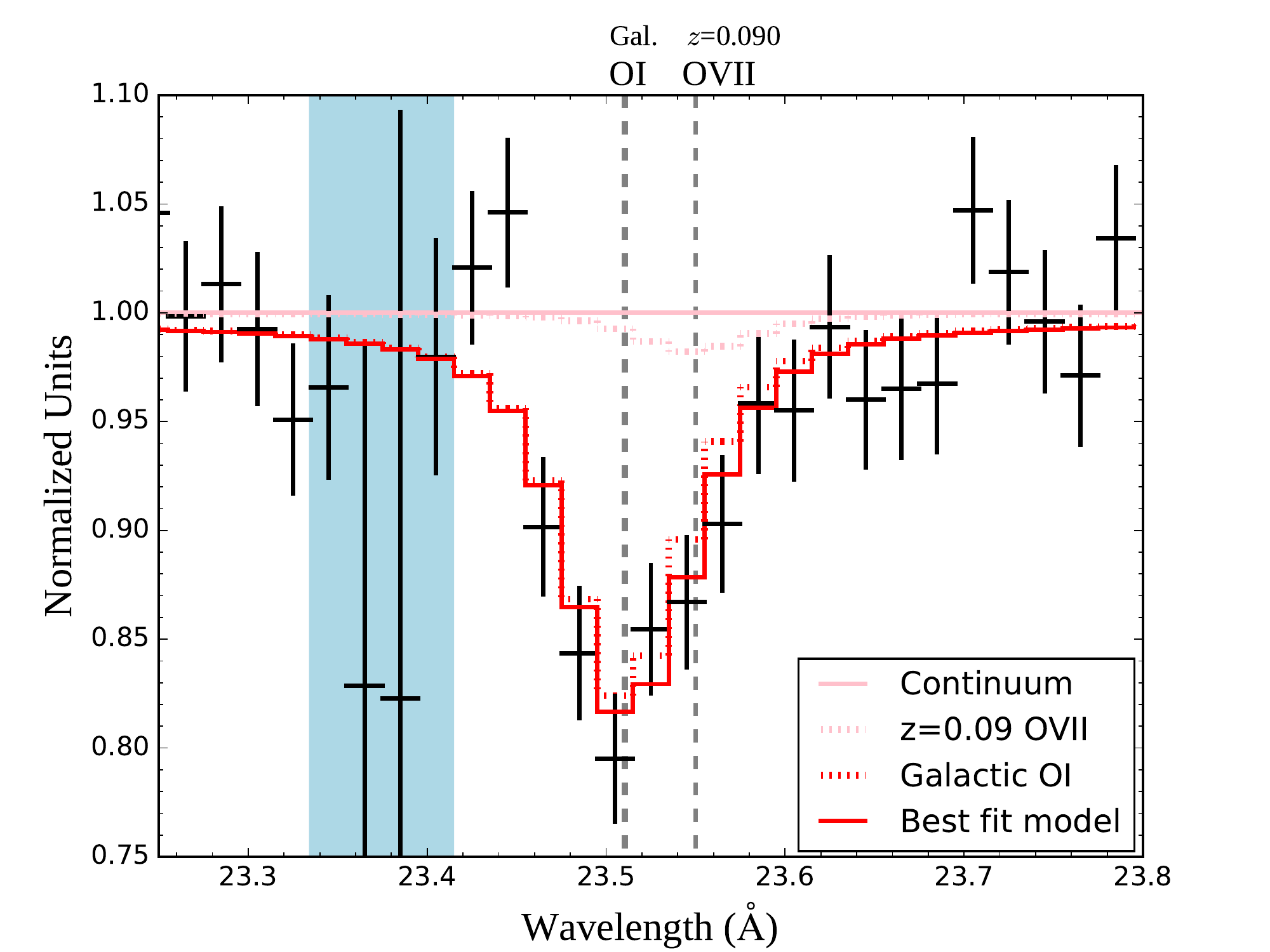}
\end{center}
\caption{Zoom-in on the continuum-normalized, RGS1 first-order co-added spectrum at the wavelength band of the Galactic \ion{O}{I} ($\lambda\approx23.51$ \AA) doublet, where the best-fit CIE WHIM model predicts an \ion{O}{VII} He$\alpha$ line ($\lambda\approx23.55$ \AA). The different model components comprising the best-fit model (red line) are shown separately. The shaded area marks bins affected by instrumental absorption.}
\label{fig:OI}
\end{figure}

In Fig.~\ref{fig:OI}, we show the best-fit CIE WHIM model prediction for the redshifted \ion{O}{VII} He$\alpha$ line, which could not be examined independently due to blending with the Galactic \ion{O}{I} line but, nevertheless, contributes in constraining the free parameters of the CIE model. It is evident from this figure that within the obtained temperature constraints, the \ion{O}{VII} He$\alpha$ would not be strong enough for detection even without the incidental line blending.

\section{Interpretation of the results}

Here we discuss physical interpretations regarding to measurement results of X-ray and FUV absorption at $z=0.09$, assuming that both FUV and X-ray detections are real. First, in Sect. \ref{filament}, we consider the possibility that the absorbers are part of a WHIM filament crossing the sight-line at $z\approx 0.09$. In Sect. \ref{cgm}, we speculate about an alternative interpretation that the absorption signals come from the Circumgalactic medium (CGM) of individual galaxies with small impact parameters to the sight-line. Then in Sect.~\ref{phases}, we discuss the thermal structure of the absorbing gas in view of the observational results.

\subsection{Galactic environment}\label{filaments_main}

We studied the SDSS galaxy distribution in the 3C~273 sight-line around the FUV absorbers at $z\approx0.09$. 3C~273 is located in the main SDSS survey region. Our galaxy distribution analysis is based on the SDSS flux-limited ($m_r<17.77$) catalog. The catalogue and data preparation are described in \citet{Tempel:14a,Tempel:17}.

We employed the Bisous model in order to detect possible galactic filaments, which are expected to harbour a large fraction of the local baryons in the form of the WHIM. The Bisous model is a marked object point process that is specifically designed to detect galactic filaments in spectroscopic galaxy distribution data, as described in \citet{Tempel:14b,Tempel:16}. The scale of the extracted filaments in the SDSS data is $\sim$1~Mpc, which corresponds to the scale of large-scale intergalactic medium filaments in simulations (based on visual inspection of simulated filaments in \citealt{Kooistra:17}, see also the discussion in \citealt{schaye01}). Thus, if a filament axis passes the sight-line at a distance closer than 1~Mpc, we consider it to be crossing the sight-line.

\begin{figure}
\begin{center}
\includegraphics[width=\columnwidth]{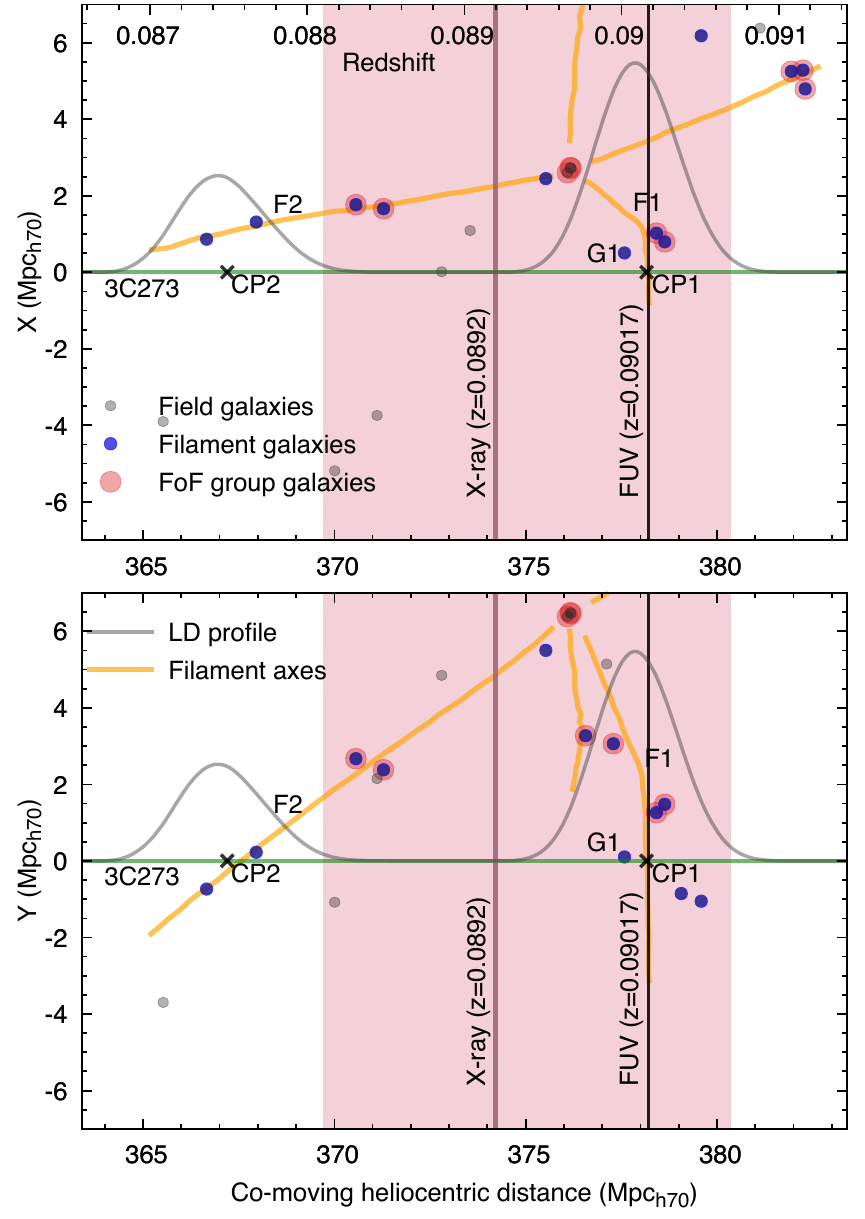}
\end{center}
\caption{Filament analysis for 3C~273 sight-line near $z=0.09$. The figure shows the distribution of SDSS galaxies (points) in two orthogonal projections (upper and lower panels), each of which are 20 Mpc thick. Blue points represent galaxies in filaments and red points indicate additional galaxies in friends-of-friends groups. The detected filament axes F1 and F2, and the associated luminosity density profiles are denoted by yellow lines and grey curves, respectively. The crossing points of the filaments F1 and F2 and the sight-line to 3C~273 are denoted with labels ``CP1'' and ``CP2'', respectively. The redshift of the \ion{O}{VI} absorption is marked with the black vertical line, whereas the pink vertical line and area correspond to the X-ray measured \ion{O}{VIII} line centroid and its 1$\sigma$ measurement uncertainty, obtained in the Gaussian model fit to the \ion{O}{VIII} Ly$\alpha$ feature (see Table \ref{table:spectral_shifts})
}
\label{fig:filaments}
\end{figure}

\subsubsection{Absorption by WHIM filaments at z=0.09}\label{filament}
We detected two filaments of $\sim$10~Mpc length at $z\approx 0.09$ around the 3C~273 sight-line, as presented in Fig. \ref{fig:filaments} (see also the discussions in \citealt{williger10}). 
Filament F1 passes the 3C~273 sight line at a 3D distance smaller than 0.5 Mpc, that is, closer than the expected filament width. Thus we consider that 3C~273 sight line crosses the core of the filament F1 at a co-moving heliocentric distance of $\approx378$~Mpc. This crossing point is consistent with the FUV and X-ray line centroids, if their redshifts are entirely due to the Hubble expansion. Filament F2 (see Fig. \ref{fig:filaments}) crosses the 3C~273 sight-line $\sim$10~Mpc away from the FUV centroid (closer to the observer), if the FUV absorber has no significant radial peculiar velocity component. Assuming alternatively that the FUV absorber is the filament F2, it should have a radial velocity of 800~$\mathrm{km}~\mathrm{s}^{-1}$. Infall with such a high velocity is unlikely, indicating that the FUV absorber is located close to the crossing point of filament F1 and the sight-line towards 3C~273 (see Fig. \ref{fig:filaments}). Given the relatively large statistical uncertainties in the redshift of the X-ray absorber, and allowing an infall velocity of a few 100~km~s$^{-1}$ (i.e., a shift of a few Mpc in location), we cannot determine whether F1 or F2 is a more likely location for the X-ray absorber. However, the consistency of the FUV and X-ray centroid redshifts with each other and with the location of the major galactic filament F1 suggests that both absorbers are due to WHIM in F1.

Using the galaxy distribution we also generated a three dimensional luminosity density (LD) field \citep[for details, see][]{Liivamagi:12,Tempel:14a}. The LD profile along the sight-line (see Fig. \ref{fig:filaments})
peaks at the crossing points of filaments F1 and F2.
Assuming that the X-ray absorber is located at F1, and following the procedures described in \citet{Nevalainen:15}, we converted the luminosity density profile along the sight-line in the radial range of 375--381~Mpc into a WHIM hydrogen column density estimate of ${N}_\mathrm{H,F1} \sim 2 \times 10^{19}~\mathrm{cm}^{-2}$. (Currently we cannot explicitly account for the possible selection effect induced uncertainties in the total error budget of $N_\mathrm{H}$, and consequently we only give the best estimate value here). 
The perpendicular orientation of F1 relative to the 3C~273 sight-line provides a minimal path length ($\sim$ 1 Mpc) through the filament and thus gives a relatively low column density. Assuming a 1 Mpc path length and a constant density along the path yields a WHIM hydrogen number density
of $\sim 6 \times 10^{-6}$ cm$^{-3}$, that is, a baryon overdensity of $\sim 20$, a value consistent with the WHIM in simulations. We note that the estimate of the total hydrogen column density of F1 ($\sim 2 \times 10^{19}~\mathrm{cm}^{-2}$) is consistent with the value the X-ray data yielded for ${N}_\mathrm{H}$ (Table \ref{table:news}), if the WHIM metallicity is $ \sim 10^{-1}$ Solar (as we find that $N_\mathrm{H}^{X-ray} \gg N_\mathrm{H}^{FUV}$, see details in Sect. \ref{phases}).

\begin{figure*}[t]
\begin{center}
\begin{minipage}{0.494\textwidth}
\includegraphics[width=\textwidth]{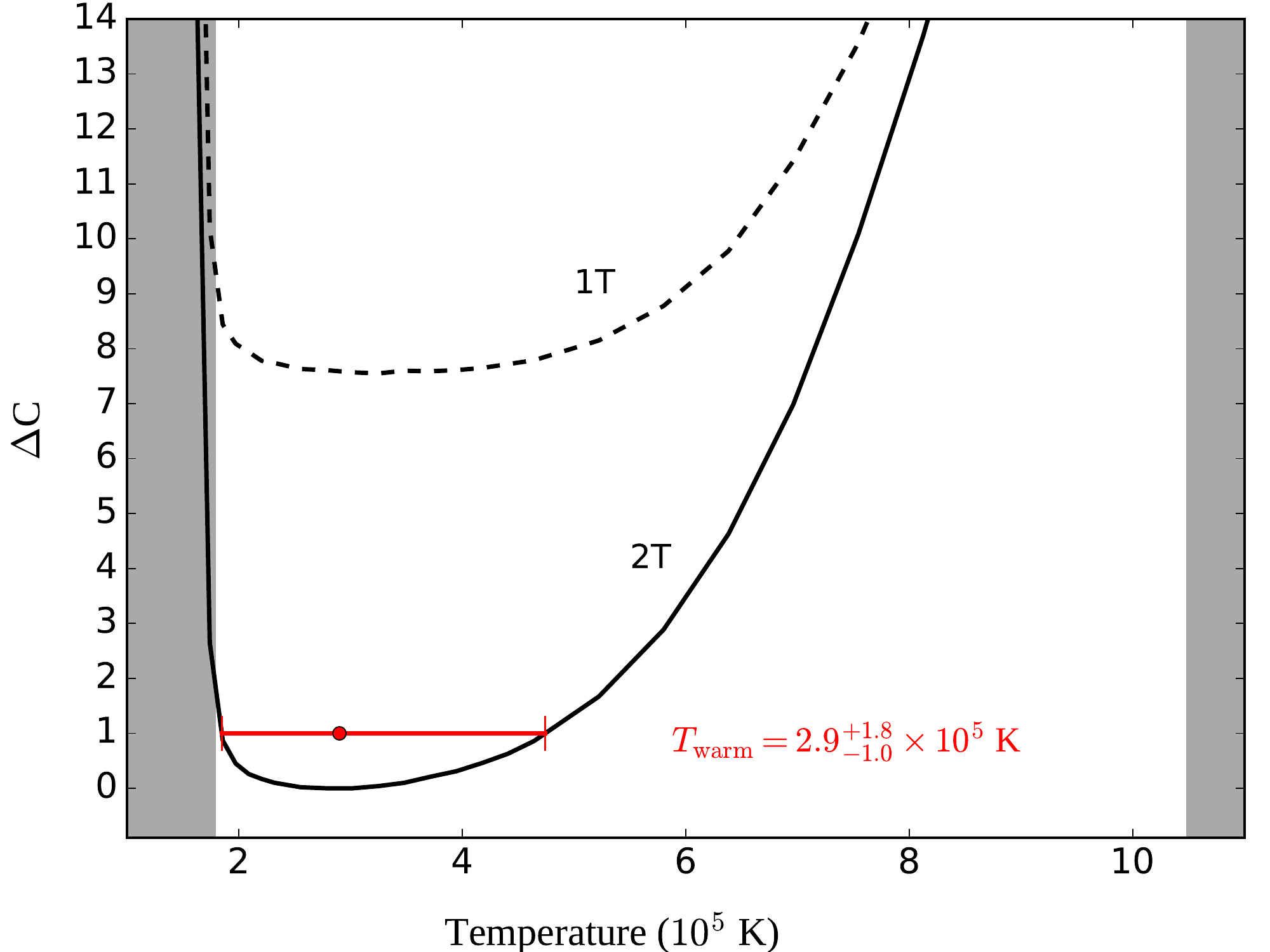}
\end{minipage}
\begin{minipage}{0.494\textwidth}
\includegraphics[width=\textwidth]{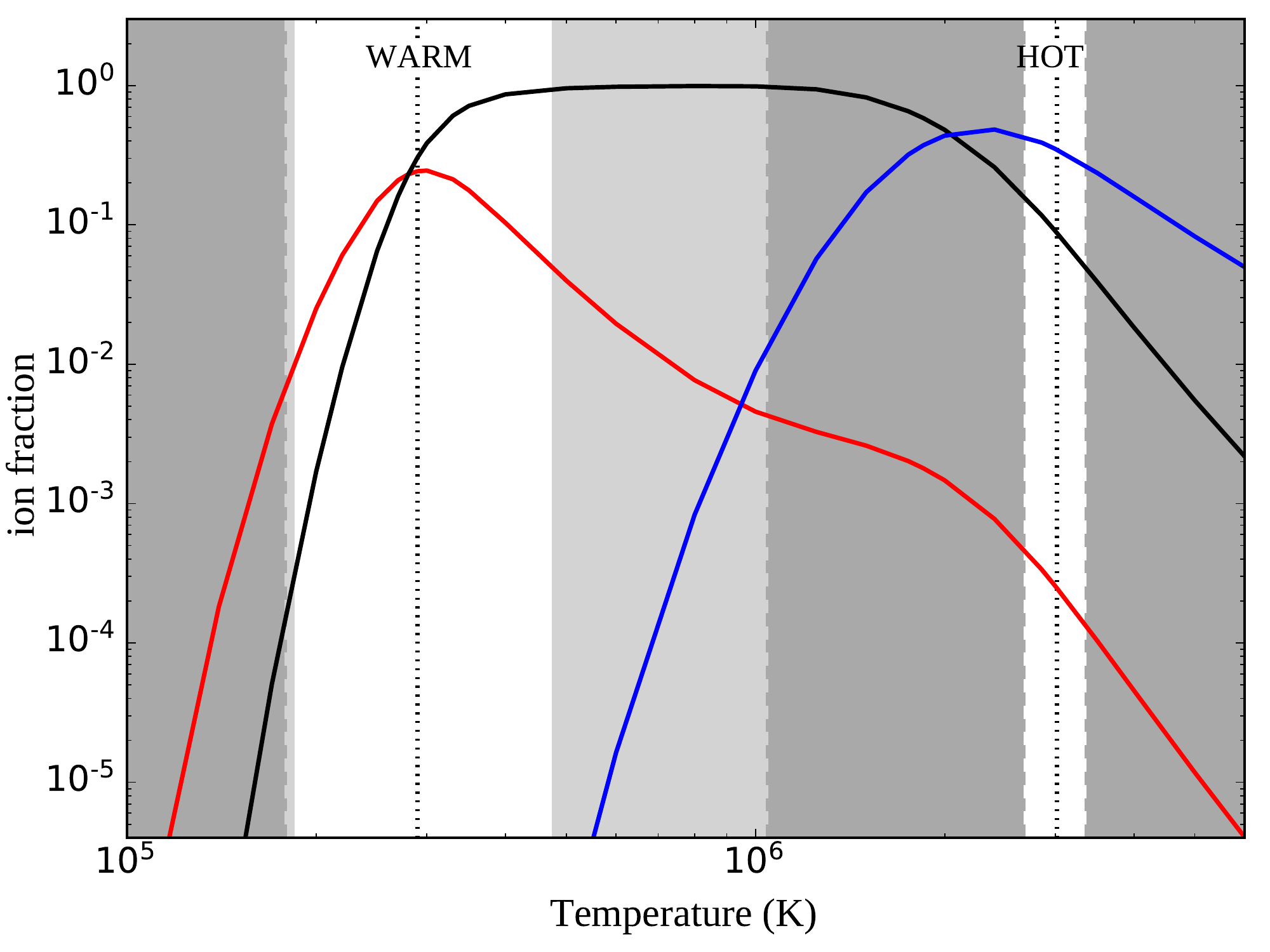}
\end{minipage}
\end{center}
\caption{Thermal properties of the absorbing gas at $z=0.09017$ assuming CIE. Left: Temperature dependence of the two-phase absorption model fit statistics to the X-ray data (black solid curve) when $N_\mathrm{OVI}$ is fixed to the FUV measured value. The shaded areas mark the forbidden regions based on the FUV measurements (the lower limit is yielded by the limits on $N_\mathrm{CIII}$ and $N_\mathrm{OVI}$, whereas the upper limit is set by the \ion{O}{VI} line broadening). The 1T curve shows the X-ray model fit statistics without the hot absorption component included, and is plotted for reference. 
Right: \ion{O}{VI} (red), \ion{O}{VII} (black) and \ion{O}{VIII} (blue) ion fractions over the $T$ -ranges of the warm and hot CIE components. The dotted lines mark the best-fit temperatures for both phases, while the white regions mark the $1\sigma$ uncertainty limits as yielded by X-ray data. The light gray shaded regions show the improvement which X-ray data provides to the $T_\mathrm{warm}$ determination as compared to the limits obtained with FUV data.}
\label{fig:interpretation}
\end{figure*}

\subsection{Absorption by Circumgalactic Medium at z=0.09}\label{cgm}

We next considered the possibility that the haloes of galaxies close to the 3C~273 sight-line are the the dominant source of the FUV and X-ray absorption we associated with the F1 filament above. 
The most likely object to provide sufficient ion columns is G1 (see Fig. \ref{fig:filaments}), the nearest galaxy to the sight-line ($b_\mathrm{impact} \approx 500$ kpc) at a redshift close to F1. It is a spiral galaxy with $M_r = -21.1$, and hence photo-metrically similar to the Milky Way \citep{2015ApJ...809...96L}.
Using the \texttt{IllustrisTNG} simulations \citep{nelson18} for galaxies with halo mass log$(M_\mathrm{halo}) = 12.0$, that is, the same as for the Milky Way, for $b_\mathrm{impact} \approx 500$ kpc we obtained an average log$N_\mathrm{OVI}(\mathrm{cm^{-2}})\approx 12.2$ with 1 $\sigma$ halo-to-halo variation of $\approx 9.8-13.0$ (D.~Nelson priv. comm.). Therefore the \ion{O}{VI} column of the $z=0.09$ absorber, log($N_\mathrm{OVI})=13.263 \pm 0.110$ \citep{tilton12}, seems consistent with absorption through the halo of G1. However, given that the halo-to-halo variation spans over three order of magnitudes, the consistency may be only coincidental and one cannot determine the origin of the absorption without further information on the absorber properties.

When it comes to the hot phase, we do not have similar simulation results available, and the relevant observational information is also very limited. As it stands, hot ($\sim 10^6$ K) coronae around spiral galaxies have only been directly measured 
for a few nearby massive spirals, and even in these cases the photon statistics have limited the detailed analysis to
the centermost halo regions. For instance, \citet{bogdan13} reported hot haloes around the NGC~1961 and NGC~6753 spirals, in which $\approx 50-70$ \% of baryons were found to be missing from the volumes enclosed within $R_\mathrm{vir}$. These results were later confirmed by \citet{andersson16} and \citet{bogdan17} with use of deeper data.

We note however, that there are indications that hot galactic haloes may extend very far out from the host galaxies \citep[e.g.,][]{wakker09,tumlinson11,stocke12,johnson15,burchett19}, up to several hundred kpc distances in Milky Way-like galaxies (see e.g., \citealt{gupta12}; \citealt{fang13}). 
In the case of 3C~273, the sight-line has a $\sim 2.5 R_\mathrm{vir}$ impact parameter with respect to the galaxy G1. Hence, linking the X-ray absorption to the hot halo of G1 would require a substantial CGM, extending far beyond the virial radius. Unfortunately, the available X-ray instrumentation is not capable of providing the information required
to determine the mass distribution of hot gas around spirals out to very large radii.
(Note that current measurements of the missing baryon budget in spirals, such as those quoted above, are based on parametric profiles, extrapolated from a small fraction of $R_\mathrm{vir}$ around the central parts of the halo.) Therefore, while we cannot confirm or exclude the possibility that the 3C~273 hot gas absorption would mainly be associated with galaxy G1 (rather than with the filamentary WHIM gas residing in structure F1), this interpretation seems unlikely.

\subsection{WHIM gas phases at z=0.09}\label{phases}

In this work, we have shown that X-ray grating data yields evidence of hot absorbing gas at $z\approx0.09$. \cite{tilton12}, building on the work of \cite{danforth08}, \cite{tripp08}, and \cite{sembach01} reported FUV detections of \ion{O}{VI} $\lambda\lambda$~1031.9, 1037.6~\AA\, lines at the corresponding $z$ at 5.0 and 2.8~$\sigma$ significance levels, indicating a log$N_{\mathrm{OVI}}(\mathrm{cm^{-2}})=13.263\pm0.110$ column density for \ion{O}{VI}. The CIE WHIM model presented in this work only predicts $\sim\frac{1}{10}$ the \ion{O}{VI} column density (Table \ref{table:news}) that the FUV data indicate, meaning that a one temperature CIE model is insufficient to explain all the observational results. 

At present, observational information on the thermal structures of WHIM absorbers is absent, but, for example, the \texttt{EAGLE}  \citep{schaye15} simulations indicate that many high column density WHIM absorbers are multi-temperature \citep[e.g.,][]{oppenheimer16,wijers18}. 
For such absorbers, multiband analysis may often be required to yield information about the various temperature phases that may co-exist. This is also demonstrated by the CIE WHIM modeling results of this study; despite the fact that the hot phase predicts relatively high $N_{\mathrm{OVI}}$, its absorption imprint cannot be measured in the FUV because of the shallow line profile due to thermal line broadening. 

In the absence of detailed knowledge of the $z=0.09$ absorber's local radiation field, we do not know the precise balance of ionization processes at work. However, if the ionizing radiation field can be predominantly characterized by the typical metagalactic ultraviolet background (UVB; e.g., \citealt{HaardtMadau2012}), it is unlikely that photoionization is the primary contributor to the observed \ion{O}{VI} column densities, as photoionization calculations using a variety of UVB models fail to produce significant \ion{O}{VI} column densities \citep[e.g.,][]{tepper11, shull15,rahmati16}. Even in the circumgalactic medium of star-forming $L_*$ galaxies, photoionization is unlikely to drive the production of \ion{O}{VI} unless the physical gas densities are extremely low (\citealt{McQuinnWerk2018}), or if there has been AGN activity within the last recombination time \citep[][]{oppenheimer13,segers17,oppenheimer18}. The ionization of the observed \ion{O}{VI} absorber is thus likely driven by collisional processes. Though the frequent association of \ion{O}{VI} absorbers with photoionized gas (e.g., lower ionization states of C, N, and Si) suggests that at least some \ion{O}{VI} absorbers arise in a non-equilibrium, multiphase gas, 
we limit our further discussion to the simplified case of collisional ionization equilibrium so that we can derive plausible temperatures for the observed gas.

Assuming the CIE conditions to be valid, the FUV observational results readily set constraints on the gas temperature of the warm absorber. Namely, combining the information of the upper limit on the \ion{C}{III} ion column density (log$N_{\mathrm{CIII}}(\mathrm{cm^{-2}})<12.590$, \citealt{tilton12}) and the lower limit on $N_\mathrm{OVI}$, one gets the lower temperature limit if the relative elemental abundances are considered known (because of the \ion{C}{III}, \ion{O}{VI} ion fraction $T$-dependencies). On the other hand, assuming pure thermal line broadening for the \ion{O}{VI} line broadening parameter ($b=22.2\pm10.8$ km$\,$s$^{-1}$, Table~\ref{table:redshifts}) directly yields an upper temperature limit for absorbing gas. 

However, we found that with the help of the X-ray data, these temperature limits can be constrained more accurately. In the two panels of Fig. \ref{fig:interpretation} we present the results obtained with a two phase (warm-hot) CIE absorber model constructed around the relevant information obtained in the X-ray and FUV measurements. In the 2T model (the CIE WHIM model + additional \texttt{\small SPEX} `hot'-component for the warm absorber), the hot CIE absorption component was fixed to the best-fit values of the CIE WHIM (Table \ref{table:news}), while the total $N_{\mathrm{OVI}}$ (=$N_{\mathrm{OVI}}^{\mathrm{warm}}+N_{\mathrm{OVI}}^{\mathrm{hot}}$) was fixed to the FUV measured value of log$N_{\mathrm{OVI}}(\mathrm{cm^{-2}})=13.263$. We then examined the goodness of fit as a function of warm CIE absorber temperature. We also conducted the same study by omitting the hot WHIM component from the model, to investigate the dependence of the results on the hot CIE component (1T, Fig. \ref{fig:interpretation}). We note that in both the 1 and 2T fits, the only free parameter was the warm WHIM temperature $T_\mathrm{warm}$ (in addition to the absorbed emission model, Sect. \ref{modeling}), as $N_\mathrm{H}^{\mathrm{warm}}$ is defined by the CIE constraints from the fixed $N_{\mathrm{OVI}}$.

We found that the 2T (and 1T) model yields a global minimum C-statistic in a  temperature range matching the FUV derived $T$ limits. More precisely, the 2T fit yielded $T_\mathrm{warm}\approx2.9_{-1.0}^{+1.8}\times10^5$~K (or $kT_\mathrm{warm}\approx0.025$ keV) with $N_{\mathrm{H}}^{\mathrm{warm}}/N_{\mathrm{H}}^{\mathrm{hot}}\sim10^{-2}$ (see the left panel in Fig. \ref{fig:interpretation}). We note that this solution occurs because towards the lower temperatures, the ratio of $N_{\mathrm{OVI}}$ to lower ionization states of oxygen quickly decreases, predicting rise of prominent \ion{O}{IV} (blend at $\lambda_\mathrm{0}\approx22.7$~\AA) and \ion{O}{V} ($\lambda_\mathrm{0}\approx22.37$~\AA) lines not present in the X-ray spectra (but observed, e.g., in Galactic halo, see \citealt{nevalainen17}). In contrast, when moving towards the higher temperatures, the $\frac{ N_{\mathrm{OVII}}}{N_{\mathrm{OVI}}}$ -ratio becomes inconsistent with the X-ray data. 
The attained 2T solution is in fact the only viable two-phase solution that can simultaneously fulfill both the FUV and X-ray constraints, which can also be understood through the visualization in the right panel of Fig.~\ref{fig:interpretation}.

Considering the thermal analysis results and the redshift match between the hot and warm absorbers together implies that the FUV and X-ray absorbers are part of the same, multi-temperature structure of intergalactic gas. The redshift match does not necessarily signify strict spatial co-location of the detected gas phases, however, because different thermal phases can occupy spatially distinct physical environments within the same structure. Indeed, recent \texttt{EAGLE} simulations indicate that WHIM structures are often composed of a variety of thermodynamical environments characterized by wide range of densities, temperatures and pressures \citep[e.g.,][]{oppenheimer16, wijers18}. Such complexity limits our ability to further examine the absorber properties, by adopting the requirement of pressure equilibrium for different phases for instance, or by means of other similar constraints.

However, combining the information from the FUV and X-ray measurements does give weak constraints on the line-of-sight turbulent velocities $v_\mathrm{turb}$ associated with the warm WHIM component. Namely, writing the definition of the line Doppler parameter $b$ for $v_\mathrm{turb}$ we have
\begin{equation}
v_\mathrm{turb}=\sqrt{\frac{b^2}{2}-\frac{kT}{m}},
\end{equation}
where $m$ is the mass of a given ion, of which we measure a rest-frame spectral linewidth $b$ at gas temperature $T$. Thus, in the case in hand, the first term inside the square root can be determined by the FUV \ion{O}{VI} measurements, whereas the second term can be obtained from the X-ray temperature constraints on the warm WHIM component. (We ignore any hot phase contribution to the FUV \ion{O}{VI} line width, since we expect the hot phase to have only $\sim\frac{1}{10}\times$ the \ion{O}{VI} column and $\sim10\times$ the temperature). Using the obtained values gives us upper limit $v_\mathrm{turb}\lesssim20$ km$\,$s$^{-1}$, or $v_\mathrm{turb}/v_\mathrm{thermal}\approx0.5\pm0.5$. In order to obtain a lower limit for $v_\mathrm{turb}$, more accurate measurements of $b$ would be required. We point out that this method could be applied to constrain $v_\mathrm{turb}$ for any (collisionally ionized) warm \ion{O}{VI} absorbers in lines-of-sight where high-resolution X-ray data with good photon statistics are available.

We will now consider the spectral properties of the BLA lines predicted by the two-phase model to check whether they contradict the non-detection of BLA lines at $z=0.09$.
\cite{richter06} found that the BLAs are detectable in STIS data if
\begin{equation}\label{tres}
\frac{N_\mathrm{HI}}{b_\mathrm{HI}}\gtrsim \frac{3\times 10^{12}}{(\mathrm{S/N)}} \mathrm{cm}^{-2}\,(\mathrm{km}\,\mathrm{s}^{-1})^{-1},
\end{equation}
where (S/N) denotes the signal-to-noise ratio per spectral resolution element at the location of the BLA. We examine the detectability of the predicted BLAs using the \cite{williger10} quoted sensitivity limit for STIS spectral data at the 3C~273 sight-line (log$\,N_\mathrm{HI}/b_\mathrm{HI}\approx10.9$ at $z\approx0.09$). \\
According to the best-fit 2T model, the two BLAs are characterized by $b_\mathrm{HI}^\mathrm{warm}\approx69$~km$\,$s$^{-1}$, $N_\mathrm{HI}^\mathrm{warm}\approx10^{11.3}\times Z_\sun/Z_\mathrm{warm}$~cm$^{-2}$ and $b_\mathrm{HI}^\mathrm{hot}\approx223$~km$\,$s$^{-1}$, $N_\mathrm{HI}^\mathrm{hot}\approx10^{12.0}\times Z_\sun/Z_\mathrm{hot}$~cm$^{-2}$ (here we have neglected the non-thermal line broadening, which would effectively decrease the line detectability).
The detection criterion in Eq. \ref{tres} may then be written as $N_\mathrm{HI}^{i}/b^{i}_\mathrm{HI}\times10^{-10.9}\gtrsim Z^{i}/Z_\sun$, which yields metallicity limits $Z_\mathrm{warm}\approx0.04\times Z_\sun$ and $Z_\mathrm{hot}\approx0.06\times Z_\sun$, above which the lines are undetectable. Therefore
neither of the phases produces detectable BLAs at the expected metallicity range of filamentary WHIM, $Z=0.1-0.4\times Z_\sun$ \citep{martizzi19}, and the non-detectability is also consistent with our weak limit on the hot phase metallicity, $Z_\mathrm{hot}\sim10^{-1}$ (see Sect. \ref{filament}).

\section{Comparison to Simulations}\label{comparison}

In Sect. \ref{filament} we showed that the FUV and X-ray absorption at $z\approx0.09$ can originate from WHIM located in a large scale filamentary structure (of the Cosmic Web). Then, in Sect. \ref{cgm} we found that the nearest detected galaxy to the absorber is unlikely to produce the measured absorption. However, it is still possible that at the location of the absorber there are galaxies fainter than the SDSS detection limit, which could produce the measured level of \ion{O}{VI}. In this case, only the X-ray lines
would originate from the intergalactic medium in the filament.

In order to gain more insight of the physics of the co-located FUV and
X-ray absorbers, we investigated the \texttt{EAGLE} cosmological, hydrodynamical simulations (\citealt{schaye15}, \citealt{crain15}, \citealt{mcalpine16}). Given the various possibilities for the
environment of the $z\approx0.09$ absorbers, we did not attempt to separate the haloes and the filaments when extracting the simulated data for this study. 
In this section, we summarize the observable properties of spatially co-located hot (\ion{O}{VII}, \ion{O}{VIII} and \ion{Ne}{IX}), and warm WHIM (\ion{O}{vi}) absorbers in \texttt{EAGLE}, and compare these predictions to our measurements.

\subsection{EAGLE simulation}

The code for the \texttt{EAGLE} simulations is a modified version of \textsc{gadget3} (last described in \citealt{springel05}). It uses a TREE-PM scheme to calculate gravitational forces. One of the modifications made to \textsc{gadget3} is the use of a different hydrodynamics solver: \texttt{EAGLE} uses a smoothed particle hydrodynamics (SPH) implementation known as \textsc{anarchy} (\citealt{schaye15}, \citealt{schaller15}). Cosmological parameters were taken from \cite{planck13}. 

Gas cooling is implemented per element as described by \citet{wiersma09a}, using the abundances of 11 elements tracked in the simulation. Ionization equilibrium with collisional processes and photo-ionization by a \cite{HaardtMadau2001} UV and X-ray background is assumed. 
Since the particle mass of the simulations is $\approx 2 \times 10^{6}\, \mathrm{M}_{\sun}$, we cannot resolve the formation of and feedback (stellar winds and supernovae) from individual stars, accretion disks, jets, etc.\ in AGN. Therefore, star formation and stellar and AGN feedback on the mass scales we can resolve are implemented using subgrid models. The feedback models have free parameters, which were chosen by calibrating them to reproduce the $z=0.1$ galaxy stellar mass function, the relation between black hole mass and galaxy mass, and reasonable galaxy sizes. We use the $100^3\,\mathrm{Mpc}^3$ simulation volume (\textsc{L100N1504}) at $z=0.1$, which was run with the standard \texttt{EAGLE} subgrid parameters (the Reference model).

We compute column densities as in \cite{wijers18}. In short, we assume the gas is in collisional and photo-ionisation equilibrium with the \cite{HaardtMadau2001} UV/X-ray background at $z=0.1$. We then use the gas temperatures, densities, and oxygen mass fractions from the simulation to compute the number of ions in long, thin columns along the line of sight. The columns are thin enough that the column densities at the values we observe are converged. Their line of sight length was chosen to be $6.25\,$comoving Mpc, which corresponds to $404 \, \mathrm{km}\, \mathrm{s}^{-1}$ (rest-frame) in velocity space, ignoring peculiar velocities, or $\Delta z = 0.0015$. When comparing temperatures for the different ions, we measure ion-weighted temperatures in the same columns. We define counterpart column densities to be column densities measured in the same columns for different ions. Since we define our simulated absorbers by these fixed boxes, our column density and counterpart determinations from \texttt{EAGLE} are only an approximation of what an observer would measure in an \texttt{EAGLE} universe. 
The redshift range in these slices does roughly match the difference between the \ion{O}{VI} and X-ray centroids, although the $z$ uncertainty is large in the X-ray. 
While we did not impose any direct filament environment selection criteria on the \texttt{EAGLE} data, we note that concentrating on the densest hot WHIM, we are approximately focusing on locations close to the filament axes, as indicated by the cosmological simulations such as the \texttt{EAGLE} and \texttt{IllustrisTNG} (see \citealt{nelson17}).

\begin{figure*}[t]
\begin{center}
\includegraphics[width=0.7\linewidth]{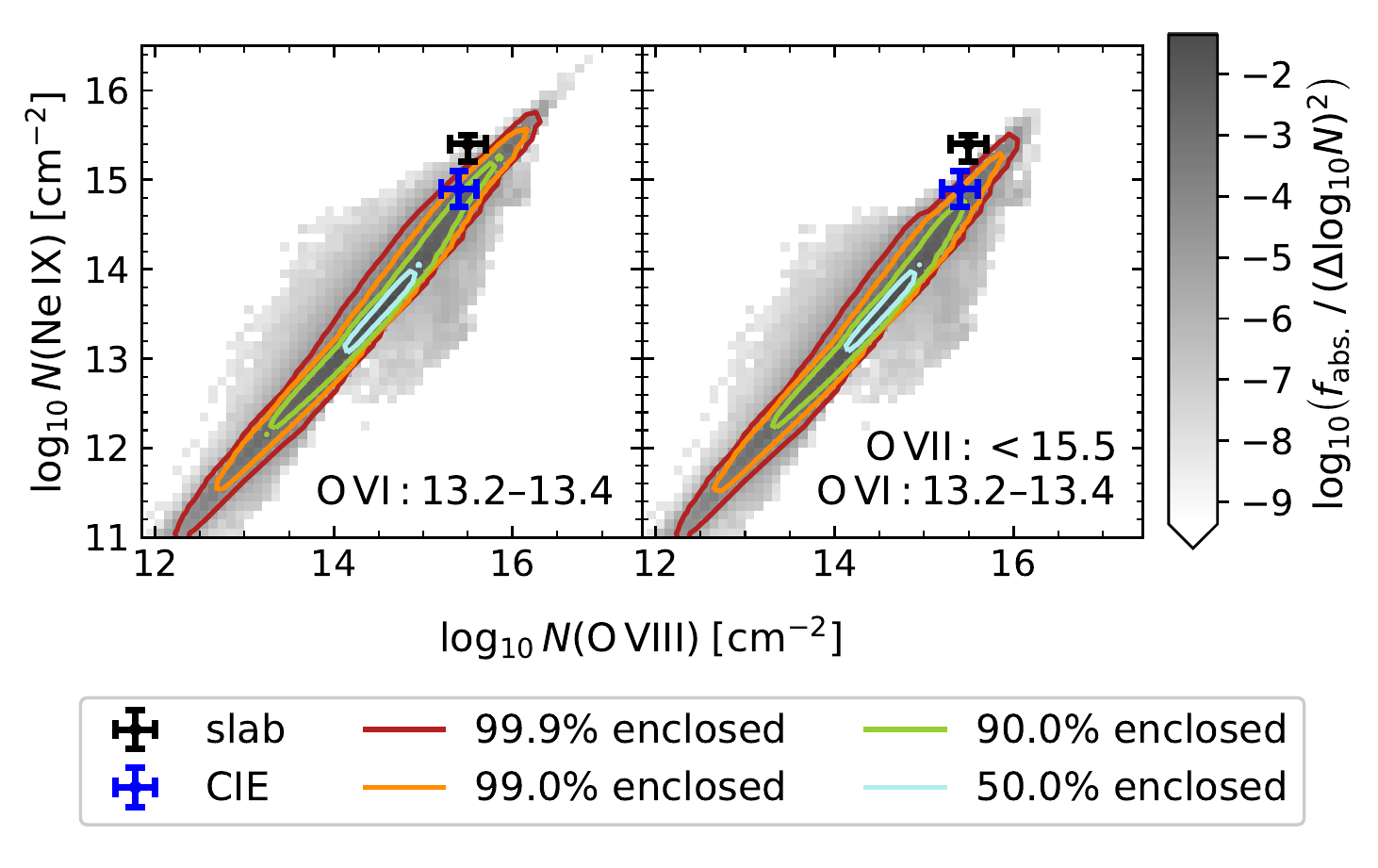}
\end{center}
\caption{The distribution of \ion{O}{VIII} and \ion{Ne}{IX} column densities in counterparts to \ion{O}{VI} absorbers (grayscale and contours) according to the \texttt{EAGLE} simulation, compared to the measured column densities at $z=0.09$ for the CIE and slab models. In the left panel we show counterparts to \ion{O}{VI} absorbers with column densities of $10^{13.2}$--$10^{13.4} \, \mathrm{cm}^{-2}$, which is roughly the $\pm 1 \, \sigma $ range for the FUV absorber of \citet{tilton12}. In the right panel, we show the subset of \ion{O}{VI} counterparts with an \ion{O}{VII} column density in agreement with the upper limit measured with the slab model. The contours enclose different fractions of the counterparts, as indicated in the legend, while the grayscale shows the fraction of all \texttt{EAGLE} absorbers satisfying the indicated constraints. Although most \texttt{EAGLE} \ion{O}{VI} absorbers have weaker (and hence undetectable) associated \ion{Ne}{IX} and \ion{O}{VIII} absorption than observed at $z=0.09$ (but consistent with the non-detection for the $z=0.12$ \ion{O}{VI} absorber), both the `slab' and CIE measuremets are consistent with \texttt{EAGLE} at the $\approx1-2\sigma$ level.
The \ion{O}{VII} upper limit does not affect the distribution much.   
}
\label{fig:Nsimcomp}
\end{figure*}

\subsection{Results}

In Fig.~\ref{fig:Nsimcomp}, we investigate what sort of X-ray counterparts we would expect to find for the \ion{O}{vi} absorber at $z=0.09$, based on its column density. In \texttt{EAGLE}, such an absorber can have a broad range of \ion{O}{VIII} and \ion{Ne}{IX} column densities, but the absorption in the two X-ray lines is strongly correlated. The measured X-ray column densities are large compared to the predictions, but not unreasonably so, especially considering the minimum detectable absorption and the non-detection of X-ray counterparts for the $z=0.12$ \ion{O}{VI} absorber. The CIE WHIM model column densities agree better with the \texttt{EAGLE} predictions than the `slab' column densities, but both are within $1-2 \, \sigma$ of the region where $90\, \%$ of absorbers consistent with the observed \ion{O}{VI} column densities lie in \texttt{EAGLE}, if the observed upper limit on \ion{O}{vii} is ignored (left panel of Fig. \ref{fig:Nsimcomp}). Including the `slab' model \ion{O}{VII} upper limit means the \ion{O}{VIII}/\ion{Ne}{IX} absorption measured with the `slab' model agrees slightly less well with \texttt{EAGLE} (right panel of Fig. \ref{fig:Nsimcomp}). 
However, given the $\sim 10^{15} \, \mathrm{cm}^{-2}$ detection limit for \ion{O}{viii} and \ion{Ne}{ix}, any of the most likely $\sim 90 \, \%$ of \ion{O}{vi}, \ion{O}{vii} upper limit counterparts would be undetectable with current instrumentation.
Considering this selection effect, and the non-detection of a hot counterpart for the \ion{O}{VI} absorber at $z=0.12$ (characterized with  similar $N_\mathrm{OVI}$ to $z=0.09$ absorber, Table \ref{table:redshifts}), we find that the FUV and X-ray ion column density measurements 
are generally consistent with the \texttt{EAGLE} predictions.

\begin{figure}[t]
\begin{center}
\includegraphics[width=\columnwidth]{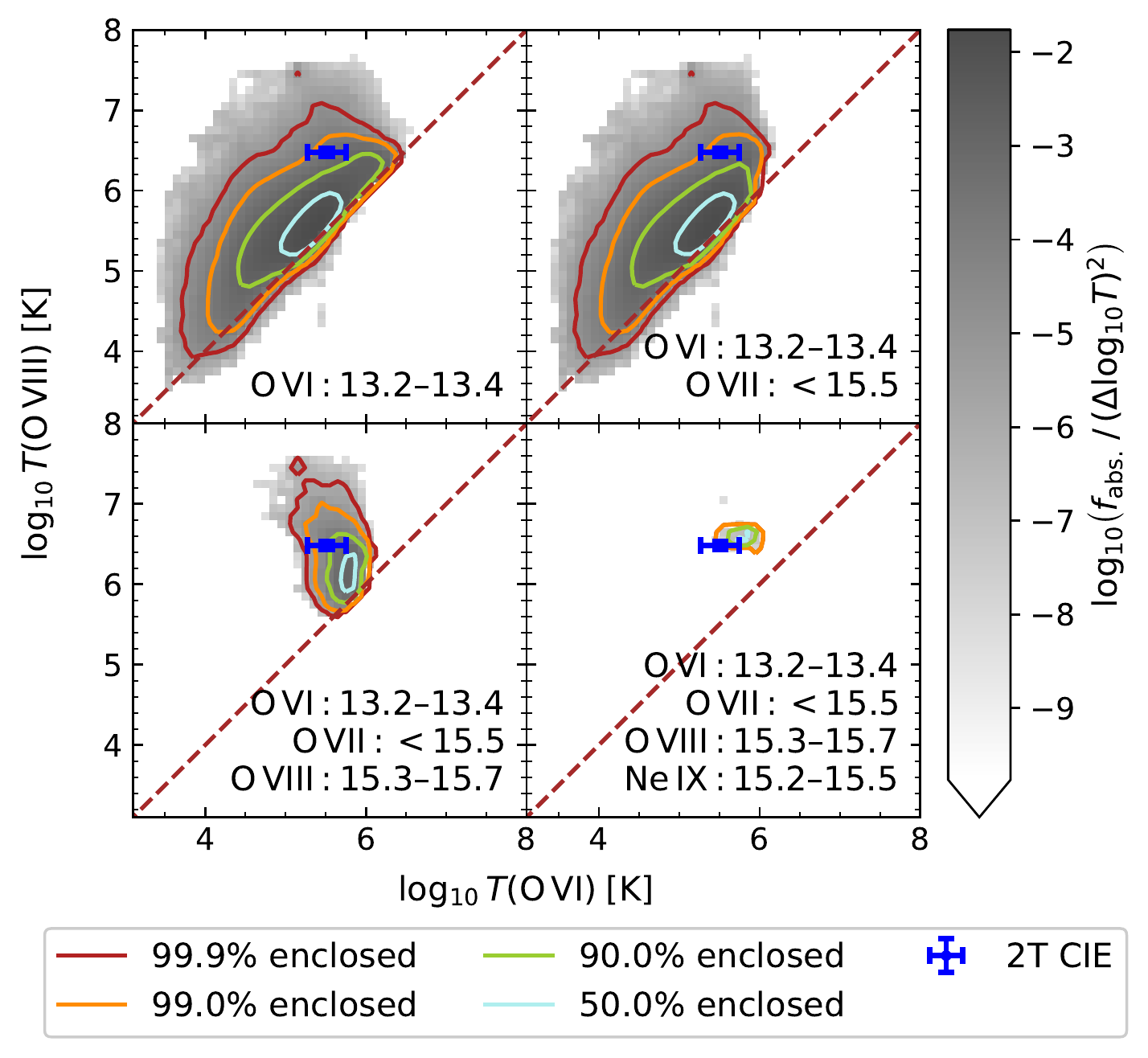}
\end{center}
\caption{The distribution of \ion{O}{VI}- and \ion{O}{VIII}-weighted temperatures in counterparts to \ion{O}{VI} absorbers (grayscale and contours) according to the \texttt{EAGLE} simulation, compared to the best-fit temperatures at $z=0.09$ for the 2-temperature CIE model (warm and hot phase, respectively). From left to right, top to bottom, we show how constraints on the column densities of different ions from the `slab' model affect the temperatures of \ion{O}{vi} and \ion{O}{viii} absorbers in \texttt{EAGLE}. Column density ranges are shown in units of $\log_{10} \, \mathrm{cm}^{-2}$ for the different ions. The \ion{O}{vi} constraints are from \citet{tilton12}. The contours enclose different fractions of the absorbers in each column density selection, as indicated in the legend, while the grayscale shows the fraction of all \texttt{EAGLE} absorbers satisfying the indicated constraints. The brown dashed line indicates where the temperatures are equal. Most absorbers are multiphase and the 2T CIE model is consistent with \texttt{EAGLE}.}
\label{fig:Tsimcomp}
\end{figure}

Then we ask the question of whether the \texttt{EAGLE} simulations support the 2-temperature solution we obtained
with the X-ray and FUV spectroscopy for the absorber at $z=0.09$. To answer this, we performed a test where we
compared our 2T-fit temperatures $T_\mathrm{warm}$ and $T_\mathrm{hot}$ (Sect. \ref{phases}) with corresponding
quantities obtained from the simulation: $T_\mathrm{OVI}$ and $T_\mathrm{OVIII}$ ion mass weighted temperatures. 
For the comparison, we had to choose a subsample of the simulations
which represents the conditions of the $z=0.09$ absorber. Since we were testing the temperatures, we obviously did not use the temperatures to select the subsample. Instead, we imposed the  measured constraints on the absorber's \ion{O}{vi} \citep{tilton12}, \ion{O}{vii}, \ion{O}{viii} and \ion{Ne}{ix} (`slab' constraints) column densities as the selection criteria when extracting the simulation data.

We illustrate the result of this study in Fig. \ref{fig:Tsimcomp}, where we examine the effect of different $N_\mathrm{ion}$ constraints on \texttt{EAGLE} distribution one by one. Adding just the \ion{O}{VI} constraint readily shows that \ion{O}{VIII} and \ion{Ne}{IX} often trace gas that is at different temperatures than the \ion{O}{VI}-rich phase (top left panel). We find that adoption of the \ion{O}{VII} upper limit does not strongly constrain the number of absorption systems in \texttt{EAGLE} (top right panel), whereas the opposite is true for \ion{O}{VIII} and \ion{Ne}{IX}, which strongly limit the temperature phase distribution of the absorption systems (bottom panels). Indeed, when all the $N_\mathrm{ion}$ measurement constraints are applied, the necessity of 2T conditions emerges in \texttt{EAGLE}. This subsample behaves well in the  $T_\mathrm{OVI}$-$T_\mathrm{OVIII}$ plane, forming a regular, relatively narrow region. Notably, we find that absorption systems with column densities like those of the $z=0.09$ absorber tend to have \ion{O}{vi} and \ion{O}{viii} temperatures that agree well with the 2T~CIE temperatures (bottom right panel). Thus, we find that the 2T CIE model is well supported by the simulations, which validates the use of CIE modeling adopted in this work.

\section{Discussion}

In this work we search for the absorption imprints of hot WHIM at the redshifts of warm intergalactic (\ion{O}{VI}) absorbers. We examine the X-ray data in a sight-line which provides one of the highest level photon statistics currently available, including observational data from 5 different high spectral-resolution X-ray instruments. Our analysis yielded a positive indication at one out of two examined \ion{O}{VI} redshifts. 
However, given that we only analyzed two \ion{O}{VI} absorbers, the uncertainty in the observational detection rate is large and one cannot draw a general conclusion from such small number statistics. Indeed, \texttt{EAGLE} predicts only an 11~\% chance of finding an \ion{O}{VIII} counterpart as strong as we found (CIE model) to at least one of the two examined \ion{O}{VI} absorbers, indicating that the observational detection rate was driven by luck.

Based on this study, we are able to make a few general remarks regarding FUV-guided searches for hot WHIM. First, two search criteria (as defined in Sect. \ref{redshifts}) were initially set in order to filter out all but the most promising FUV absorber redshifts for the X-ray analysis. In hindsight it seems, however, that the BLA criterion might have only limited use for this. Namely, considering the inevitably large $b_\mathrm{HI}$ associated with gas phases not producing a detectable \ion{O}{VI} signal in the FUV yet still producing strong lines in X-ray, such BLA detections seem unlikely.
This is because robust detections of broad HI lines require high $N_\mathrm{HI}$, in which cases the absorber metallicities should be low given the current upper limits on WHIM ionic columns, but still high enough to enable X-ray detection. Such special requirements for the environment likely mean the number of X-ray absorbers only traceable with BLAs is small.

Second, \ion{O}{vii} is often expected to be the most prominent X-ray detectable ion species to accompany \ion{O}{VI} absorbers, which is a justified assumption given the Oxygen ion fraction temperature dependencies. However, the evidence presented in this work demonstrates that \ion{O}{VI} absorbers can also be used to reveal locations of hot X-ray absorbers whose \ion{O}{vii} signal is below detectability, or in other words, when the circumstances are such that the two bands would effectively detect thermally distinct gas phases. 
As the combination of low \ion{O}{VI} ion fractions and the broad line shapes of the hot WHIM phases significantly reduces the \ion{O}{VI} FUV detectability, the FUV measured $b_\mathrm{OVI}$ should not been interpreted as a direct proxy for the (maximum) temperature of the absorber. Rather, $b_\mathrm{OVI}$ may often be merely a property of the most easily detectable gas phase within an otherwise thermally complex structure, as may also be inferred from the top panels of Fig. \ref{fig:Tsimcomp}.

\section{Conclusions}

In this work we examined the \ion{O}{VI} detection driven method to find missing baryons in the hot WHIM phase.
We analyzed the available high-resolution spectral data of XMM-\emph{Newton} RGS, \emph{Chandra} LETG and MEG for the quasar 3C~273, which is one of the brightest quasars in X-ray. 
We searched for the hot WHIM absorption signatures at the two redshifts where \ion{O}{VI} absorbers have been significantly detected in the FUV band.
Our main results include:
\begin{enumerate}

\item
At one of the FUV determined redshifts, $z=0.09017$,
spectral analysis with a line absorption model yields two X-ray line candidates whose centroid wavelengths match the \ion{Ne}{IX}~He$\alpha$ ($W_\mathrm{rest-frame}=2.7\pm0.9$ m\AA, confidence level $2.9~\sigma$) and \ion{O}{VIII}~Ly$\alpha$ ($W_\mathrm{rest-frame}=4.3\pm1.6$ m\AA, $2.6~\sigma$) lines. These two ion species are prominent in the same temperature range ($T\approx10^{6-6.5}$~K), thus yielding $3.9~\sigma$ combined (quadratically summed) confidence level for the hot phase lines. 
\\
\item Examining the spectral characteristics at the same redshift assuming collisional ionization equilibrium (CIE) for the absorber yields $kT=0.26\pm0.03$ keV and $N_\mathrm{H}=1.3_{-0.5}^{+0.6}\times10^{19} Z_\sun/Z$ cm$^{-2}$ for the absorbing medium. 
The column densities of \ion{O}{viii} and \ion{Ne}{ix} predicted by this model (log$\,N_\mathrm{\ion{O}{VIII}}(\mathrm{cm^{-2}})=15.4\pm0.2$, log$\,N_\mathrm{\ion{Ne}{IX}}(\mathrm{cm^{-2}})=14.9\pm0.2$) are in close agreement with those measured phenomenologically from the line flux (log$\,N_\mathrm{\ion{O}{VIII}}(\mathrm{cm^{-2}})=15.5\pm0.2$, log$\,N_\mathrm{\ion{Ne}{IX}}(\mathrm{cm^{-2}})=15.4_{-0.2}^{+0.1}$), indicating the hot phase is well described by the CIE model. 
\\
\item The thermal analysis implies a multi-temperature structure for the $z\approx0.09$ absorber, with dominant temperature components of $T_\mathrm{warm}\approx3\times10^5$~K and $T_\mathrm{hot}\approx3\times10^6$~K, and hydrogen column density ratio $N_\mathrm{H}^\mathrm{warm}/N_\mathrm{H}^\mathrm{hot}\sim10^{-2}$, when assuming CIE conditions and identical metallicities for both phases (Fig. \ref{fig:interpretation}). 
These results match the \texttt{EAGLE} hydrodynamical simulation temperature predictions for co-spatial WHIM absorbers characterized with ion column densities matching the FUV and X-ray measurements (lower right panel in Fig. \ref{fig:Tsimcomp}). 
\\
\item SDSS galaxy distribution modeling detected a large scale filament of galaxies crossing the sight-line at the location matching the FUV and X-ray absorbing gas (Fig. \ref{fig:filaments}). The total WHIM hydrogen column density 
at the 3C~273 sightline through the filament was estimated to be $N_\mathrm{H}\sim 2 \times 10^{19}$ cm$^{-2}$. Associating this value with the results of the X-ray analysis yields an absorber metallicity $Z\sim10^{-1}\times Z_\sun$, which is in the range expected for the WHIM. Combining our results hence supports the hypothesis that the X-ray and FUV absorption signals relate to a filamentary WHIM structure intersecting the 3C~273 sight-line at $z\approx0.09$.
\\
\item We found no indications of X-ray absorbing gas at the other of the two examined \ion{O}{VI} redshifts, $z=0.12005$. Since this absorber is characterized by $N_\mathrm{OVI}$ similar to the absorber at $z=0.09017$, we have a nominal 50 \% (1 of 2) detection-rate for X-ray absorbing counterparts to $N_\mathrm{OVI}(\mathrm{cm^{-2}})\approx13.2-13.4$ FUV absorbers at the examined sight-line. 
By comparison, \texttt{EAGLE} predicts $11$~\% likelihood to find at least one log$N_\mathrm{OVIII}(\mathrm{cm^{-2}})\geq15.4$ counterpart per two randomly selected \ion{O}{VI} absorbers in the coincident $N_\mathrm{OVI}$ range.
However, regarding future observations with X-ray instruments capable to detect log$N_\mathrm{OVIII}(\mathrm{cm^{-2}})\geq15.0$ signals, such as the Athena X-IFU, the same likelihood is already $29~\%$. This reflects the potential in utilizing the information on \ion{O}{VI} absorbers in finding missing baryons in hotter, X-ray absorbing phase. 

\end{enumerate}

\begin{acknowledgements}
JA and FA acknowledge the support received under X-IFU Athena project funding of University of Helsinki. ET was supported by ETAg grants IUT40-2, IUT26-2 and by EU through the ERDF CoE grant TK133 and MOBTP86. JA and FA would like to thank F. Nicastro for valuable discussions regarding to this work.
\end{acknowledgements}

\bibliographystyle{aa} 
\bibliography{references}

\end{document}